\documentclass{emulateapj}
\usepackage{epsfig}
\usepackage{graphics}

\newlength{\colwidth}
\newcommand{\HI}{\ion{H}{1}} 
\newcommand{\HeII}{\ion{He}{2}} 
\newcommand{\FeII}{\ion{Fe}{2}} 
\newcommand{\CIII}{\ion{C}{3}} 
\newcommand{\CII}{\ion{C}{2}} 
\newcommand{\CIV}{\ion{C}{4}} 
\newcommand{\SiIII}{\ion{Si}{3}} 
\newcommand{\SiII}{\ion{Si}{2}} 
\newcommand{\SiIV}{\ion{Si}{4}} 
\newcommand{\NV}{\ion{N}{5}} 
\newcommand{\OVI}{\ion{O}{6}} 

\newcommand{\tsiiv}{\tau_{\rm SiIV}}
\newcommand{\tciv}{\tau_{\rm CIV}}
\newcommand{\thi}{\tau_{\rm HI}}
\newcommand{\tsiiii}{\tau_{\rm SiIII}}

\lefthead{Aguirre et al.}
\righthead{Silicon in the IGM}

\def\gsim{\;\rlap{\lower 2.5pt
 \hbox{$\sim$}}\raise 1.5pt\hbox{$>$}\;}
\def\lsim{\;\rlap{\lower 2.5pt
   \hbox{$\sim$}}\raise 1.5pt\hbox{$<$}\;}
\def\msol{{\rm\,M_\odot}}

\def\kms{\rm\,km\,s^{-1}}
\def\mpc{{\rm\,Mpc}}

\def\cm{{\rm\,cm}}

\def\kms{{\rm\,km\,s^{-1}}}

\def\spose#1{\hbox to 0pt{#1\hss}}
\def\lta{\mathrel{\spose{\lower 3pt\hbox{$\mathchar''218$}}
     \raise 2.0pt\hbox{$\mathchar''13C$}}}
\def\gta{\mathrel{\spose{\lower 3pt\hbox{$\mathchar''218$}}
     \raise 2.0pt\hbox{$\mathchar''13E$}}}

\shorttitle{Silicon in the IGM}
\shortauthors{Aguirre et al.}

\begin{document}
	
\title{Metallicity of the intergalactic medium using pixel
statistics:\\ III. Silicon.\altaffilmark{1}} 
\altaffiltext{1}{Based on
public data obtained from the ESO archive of observations from the UVES
spectrograph at the VLT, Paranal, Chile and on data obtained at the W. M. Keck 
Observatory, which is operated as a scientific partnership among the
California Institute of Technology, the University of California, and
the National Aeronautics and Space Administration. The W. M. Keck
Observatory was made possible by the generous financial support of the
W. M. Keck Foundation.}

\author{Anthony Aguirre\altaffilmark{2},Joop~Schaye\altaffilmark{3}, \\
Tae-Sun Kim\altaffilmark{4,5}, Tom Theuns\altaffilmark{6,7}, Michael
Rauch\altaffilmark{8}, Wallace L. W. Sargent\altaffilmark{9}}
\altaffiltext{2}{Department of Physics, University of California at
  Santa Cruz,1156 High Street,Santa Cruz, CA  95064; aguirre@scipp.ucsc.edu} 
\altaffiltext{3}{School of Natural Sciences, Institute for Advanced
Study, Einstein Drive, Princeton NJ 08540; schaye@ias.edu}
\altaffiltext{4}{European Southern Observatory,
Karl-Schwarzschild-Strasse 2, D-85748 Garching bei M\"unchen, Germany}
\altaffiltext{5}{Institute of Astronomy, Madingley Road, Cambridge CB3
0HA, UK} 
\altaffiltext{6}{Institute for Computational Cosmology, Department of
Physics, University of Durham, South Road, Durham, DH1 3LE, UK}
\altaffiltext{7}{University of Antwerp, Universiteits plein 1,
B-2610 Antwerpen, Belgium}
\altaffiltext{8}{Carnegie Observatories, 813 Santa Barbara Street,
Pasadena, CA 91101} 
\altaffiltext{9}{Department of Astronomy, California Institute of
Technology, Pasadena, CA 91125}  

\setcounter{footnote}{0}

\begin{abstract}

We study the abundance of silicon in the intergalactic medium by
analyzing the statistics of \SiIV, \CIV, and \HI\ pixel optical depths
in a sample of 19 high-quality quasar absorption spectra, which we
compare with realistic spectra drawn from a hydrodynamical simulation.
Simulations with a constant and uniform Si/C ratio, a C distribution
as derived in Paper II of this series, and a UV background (UVB) model
from Haardt \& Madau reproduce the observed trends in the ratio
of \SiIV\ and \CIV\ optical depths, $\tsiiv/\tciv$.  The ratio
$\tsiiv/\tciv$ depends strongly on $\tciv$, but it is nearly
independent of redshift for fixed $\tciv$, and is inconsistent with a
sharp change in the hardness of the UVB at $z \approx 3$.  Scaling the
simulated optical depth ratios gives a measurement of the global Si/C
ratio (using our fiducial UVB, which includes both galaxy and quasar
contributions) of [Si/C]$=0.77\pm0.05$, with a possible systematic
error of $\sim 0.1\,$dex.  The inferred [Si/C] depends on the shape
of the UVB (harder backgrounds leading to higher [Si/C]), ranging from
[Si/C]$\simeq 1.5$ for a quasar-only UVB, to [Si/C]$\simeq 0.25$ for a
UVB including both galaxies and artificial softening; this provides
the dominant uncertainty in the overall [Si/C]. Examination of the
full $\tsiiv/\tciv$ distribution yields no evidence for inhomogeneity
in [Si/C] and constrains the width of a lognormal probability distribution
in [Si/C] to be much smaller than that of [C/H]; this implies a
common origin for Si and C. Since the inferred [Si/C] depends on the
UVB shape, this also suggests that inhomogeneities in the hardness of
the UVB are small. There is no evidence for evolution in [Si/C].
Variation in the inferred [Si/C] with density depends on the UVB and
rules out the quasar-only model unless [Si/C] increases sharply at low
density.  Comparisons with low-metallicity halo stars and
nucleosynthetic yields suggest either that our fiducial UVB is too
hard or that supermassive Population III stars might have to be included.
The inferred [Si/C], if extrapolated to low density, corresponds to a
contribution to the cosmic Si abundance of [Si/H]$=-2.0$, or
$\Omega_{\rm Si} \simeq 3.2\times10^{-7}$, a significant fraction of
all Si production expected by $z \approx 3$.

\end{abstract}

\section{Introduction}
\label{sec-intro}
Observational studies using high-resolution quasar absorption spectra
have firmly established the presence of heavy elements such as carbon
\citep{1995AJ....109.1522C}, silicon \citep{1996AJ....112..335S}, and
oxygen \citep{2000ApJ...541L...1S} in the diffuse intergalactic medium
(IGM). These metals constitute an important record of star formation
and of the feedback of galactic matter into the IGM.

This paper is the third in a series employing the statistics of pixel
optical depths to study the enrichment of the IGM.  The basic
technique -- pioneered by~\citet[see also Dav\'e et al. 1998 and
Songaila 1998]{1998Natur.394...44C} and later employed
by~\citet{1999ApJ...520..456E,2000AJ....120.1175E} and
\citet{2000ApJ...541L...1S} -- was developed and extensively tested
using cosmological hydrodynamical simulations by Aguirre, Schaye \&
Theuns (2001; hereafter Paper I).  The technique was then generalized
and applied to 19 high-quality spectra in order to measure the full
distribution of carbon as a function of redshift and gas density in
Schaye et al. (2003; hereafter Paper II).  See~\citet{aracil}
and~\citet{pieri} for other recent studies applying the pixel method
to \CIV\ and \OVI.

Most studies of the enrichment of the IGM have focused on \CIV\
absorption because it is strong and lies redward of the Ly$\alpha$
forest.  Moreover, as shown in Paper II, ratios of \CIV\ and \HI\
optical depths can be converted into carbon abundances using an
ionization correction that is neither very large, nor very sensitive
to the temperature and density of the absorbing gas. While
measurements of the distribution of carbon provide important
information on the mechanism by which the IGM was enriched, {\em
relative} abundance information is crucial for identifying the types
of sources responsible for the enrichment.

Previous studies have established the presence of \SiIV\ absorption in
the IGM at $z \sim 2-5$ and have used simple ionization models to
infer that their data is consistent with Si/C exceeding the solar
ratio by a factor of a few (\citealt{1996AJ....112..335S};~
\citealt{1998AJ....115.2184S};~\citealt{2001ApJ...561L.153S};
Boksenberg, Sargent \& Rauch 2003).
The observed ratios of \SiIV/\CIV\ have also been used to study the
shape and evolution of the ionizing UV background (UVB), but with
conflicting results: while~\citet{1998AJ....115.2184S} sees an abrupt
change in \SiIV/\CIV\ column density ratios at $z\simeq 3$,
\citet*{boksen} see no evidence for any evolution.

This paper presents measurements of the relative abundances of silicon
and carbon in the IGM, obtained by comparing the statistics of \SiIV,
\CIV, and \HI\ pixel optical depths in a sample of 19 high-quality
quasar spectra to synthetic spectra obtained using a cosmological,
hydrodynamical simulation. Our primary goal is to measure the overall
Si/C abundance ratio in the gas for which \SiIV\ absorption is
detected, given a model for the extragalactic ionizing UVB.  In doing
so, we also obtain some information on how much the distribution of
silicon may differ from that of carbon (beyond an overall difference
in the normalization), as well as constraints on the shape and the
evolution of the UVB and on the thermal state of the absorbing gas.

We have organized this paper as follows.  In \S\S\ref{sec-data}
and~\ref{sec-meth} we briefly describe our sample of QSO spectra and
our methodology (both of which are discussed at length in Paper II).
In \S\ref{sec-resrel} we first show results for our best QSO spectrum,
Q1422+230, as an illustration of the method.  We then give
measurements (using the full sample) of the $\tsiiv/\tciv$ ratio, from
which we infer the relative abundance of silicon to carbon ([Si/C])
for our fiducial UVB model (which is, as in Paper II, a re-normalized
version of that given for galaxies and quasars by Haardt \& Madau
2001, hereafter HM01). 
Next we give results for other UVB models, and discuss to what extent
our data can constrain the UVB shape and evolution, and the variations in
[Si/C].  We then give and interpret measurements of $\tsiiv/\thi$ and
$\tsiiii/\tsiiv$ (which help constrain the thermal state of the
absorbing gas).  We discuss and interpret our results in
\S\ref{sec-discuss}, and conclude in \S\ref{sec-conc}.

\section{Observations}
\label{sec-data}

We analyze a sample of 19 high-quality ($6.6\kms$ velocity resolution,
signal-to-noise ratio [S/N] $> 40$) absorption spectra of $2.1 \le z
\le 4.6$ quasars.  The sample is identical to that of Paper II. It
includes fourteen spectra taken with the UV-Visual Echelle
Spectrograph~\citep[UVES,][]{2000SPIE.4005..121D} on the Very Large
Telescope (VLT) and five taken with the High Resolution Echelle
Spectrograph~\citep[HIRES,][]{1994SPIE.2198..362V} on the Keck
telescope. See Paper II (\S~2) for a description of the sample and
data reduction.

\section{Method}
\label{sec-meth}

The basic method employed is similar to that described in Papers I and II, to
which we refer the reader for details and tests. Section \ref{sec-overview}
contains a brief overview of the method. Readers who want more detail
would benefit from also reading sections \ref{sec-recovery}--
\ref{sec-compmeth} which summarize the method, making reference to the
relevant sections of Papers I and II, and noting the changes to the
method used in Paper II.

\subsection{Overview}
\label{sec-overview}
For each QSO spectrum, we first recover the optical depths due to \HI\
Ly$\alpha$ ($\lambda=1216$~\AA) absorption in all pixels in the
Ly$\alpha$ forest region. Then we recover the
optical depth in pixels at the corresponding wavelengths of metal
lines such as \SiIV\ ($\lambda\lambda1394,1403$), \SiIII\
($\lambda1207$), and \CIV\ ($\lambda\lambda1548,1551$) and correlate
the (apparent) optical depth in one transition with that in another. We do
this by binning the pixels in terms of the optical depth of the most
easily detected transition, e.g., \CIV, and plotting the median optical
depth of the other transition, e.g., \SiIV, against it. A
correlation then indicates a detection of \SiIV\ absorption; see the bottom
curve in the left-hand panel of Fig.~\ref{fig:sivscat} for an
example. By doing the same for percentiles other than the 50th
(i.e., the median) we obtain information on the full probability
distribution of pixel optical depths (see the other curves in the
panel). In this way a large quantity of information
can be extracted from each observed spectrum.

By comparing the observed optical depths with those obtained from
synthetic spectra generated using a hydrodynamical simulation, we make
inferences about the distribution of silicon. For each observed
spectrum, we generate a large set of spectra drawn from the
simulation and process them to give them the same noise properties,
wavelength coverage, resolution, etc., as the observed spectrum.  We
do this for each of several UVB models; for each model the simulations
include a carbon distribution as measured in Paper II for that assumed
UVB, and some value of [Si/C],\footnote{All abundances are given by
number relative to hydrogen, and solar abundance are taken to be
$({\rm Si/H})_\odot = -4.45$ and $({\rm C/H})_\odot = -3.45$
\citep{1989GeCoA..53..197A}.}  with ionization balances computed using
the CLOUDY package\footnote{See
\texttt{http://www.nublado.org}.} (version 94; see
Ferland et al.\ 1998 and Ferland 2000 for details). We analyze these
simulated spectra in exactly the same manner as the observed ones, and
compare the simulated and observed pixel optical depth statistics.

The UVB models we employ are those in Paper II (\S~4.2).  All are based
on the models of Haardt \& Madau (2001; see also 1996)\footnote{The
data and a description of the input parameters can be found at
\texttt{http://pitto.mib.infn.it/$\sim$haardt/refmodel.html}.}, but
renormalized (by a redshift-dependent factor) so that the simulated
spectra match our measurement (Paper II) of the evolution of the mean
Ly$\alpha$ absorption. Our fiducial model, ``QG'', includes
contributions from both galaxies (with a 10\% escape fraction for
ionizing photons) and quasars; ``Q'' includes only quasars; ``QGS'' is
an artificially softened version of QG: its flux has been reduced by a
factor of ten above 4~Ryd. Model ``QGS3.2'' is like model QG for $z <
3.2$ and like model QGS for $z \ge 3.2$ and was constructed to crudely
model the possible evolution of the UVB if \HeII\ was suddenly
reionized at $z = 3.2$.

\subsection{Recovery of pixel optical depths}
\label{sec-recovery}

After continuum fitting the spectra (Paper II, \S5.1, step 1), we
derive \HI\ Ly$\alpha$ optical depths $\thi$ for each pixel between
the quasar's Ly$\alpha$ and Ly$\beta$ emission wavelengths, although
we exclude regions close to the quasar to avoid proximity effects
(Paper II, \S2). We use higher-order Lyman lines to estimate $\thi$ if
Ly$\alpha$ is saturated (i.e., $F(\lambda) < 3\sigma(\lambda)$, where
$F$ and $\sigma$ are the flux and noise arrays, see Paper I, \S4.1;
Paper II, \S5.1, step 2). For each \HI\ pixel we then derive the
corresponding \SiIV\ (and \CIV) optical depths $\tsiiv$ (and $\tciv$),
correcting for self-contamination and removing contamination by other
lines (Paper I, \S4.2).  In addition, we recover \SiIII\ optical
depths $\tsiiii$, correcting for contamination by higher-order \HI\
lines (Paper I, \S4.2). We thus have four sets of corresponding pixel
optical depths that may be correlated with each other.

\subsection{Generation of simulated spectra}

This study makes extensive comparisons between our set of observed
spectra and those generated from a cosmological simulation. The
same simulation was used in Papers I (\S3) and II (\S4.1), to which the reader
is referred for details. Briefly, the simulation uses a smoothed
particle hydrodynamics code to model the evolution of a periodic, cubic region
of a $(\Omega_m, \Omega_\Lambda, \Omega_bh^2, h, \sigma_8, n, Y) =
(0.3, 0.7, 0.019, 0.65, 0.9, 1.0, 0.24)$ universe of comoving size
$12\,h^{-1}~\mpc$, down to redshift $z = 1.5$ using $256^3$
particles for both the cold dark matter and the baryonic
components. The UVB used in the simulation was chosen to match (and
only affects) the IGM temperature as measured by~\citet{2000MNRAS.318..817S}.

Synthetic spectra are generated as described in Papers I (\S3) and II
(\S4.1), by computing the ionization balance for each particle using
an assumed uniform UVB, then passing random sightlines through the
snapshots of the simulation box and patching these sightlines together
to form one long spectrum.  Transitions due to \CIII, \CIV, \NV,
\SiIII, \SiIV, \OVI, \FeII, and 31 Lyman lines of \HI\ are included.
The spectra also include noise, instrumental broadening and
pixelization chosen to match the observed spectra in detail.

\subsection{Comparison of simulations and observations}
\label{sec-compmeth}

\begin{figure}
\epsscale{1.0} \plotone{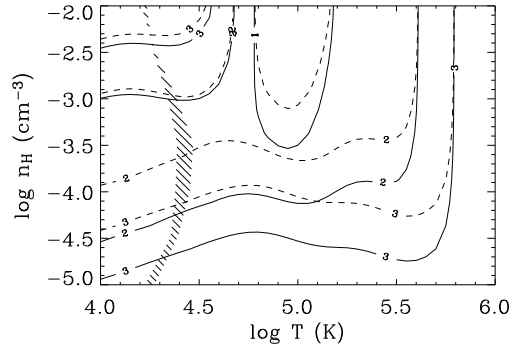} \figcaption[]{ Ionization
correction factor, $[{\rm Si/H}] - \log(\tau_{\rm SiIV}/\thi)$, as a
function of temperature and hydrogen number density. Solid
(dashed) contours are for the UVB model QG (Q) at $z=3$.
The hatched region indicates the temperature range containing 90\% of
the particles at the given density.  For both backgrounds, the
ionization correction increases rapidly for $n_{\rm H} < 10^{-4}\,{\rm
cm^{-3}}$.
\label{fig:sivhipred}}
\end{figure}

Our simulations (Paper I, \S5.1) and those of
others'~\citep[e.g.,][]{1997ApJ...488..532C,1999MNRAS.310...57S} show
that there is a tight correlation between $\thi$ and the density and
temperature of gas giving rise to the absorption.  In Paper II we used
this predicted correlation to compute an ionization correction (i.e.,
the ratios of \CIV/C and \HI/H) in order to convert ratios of
$\tciv/\thi$ into carbon abundances as a function of density (Paper I,
\S6; Paper II, \S5.1, step 5 and \S5.2).  This method works well for
\CIV\ because the ionization correction varies slowly with density
(and therefore with $\tau_{\rm HI}$) in the density range probed by
the Ly$\alpha$ forest ($n_{\rm H} \sim 10^{-6} -
10^{-3}~\cm^{-3}$). Unfortunately, this is {\em not} the case for
\SiIV: Fig.~\ref{fig:sivhipred} shows that while relatively
insensitive to gas temperature, the ionization correction factor,
$[{\rm Si/H}] - \log(\tsiiv/\thi)$, increases dramatically with
decreasing density, from $\la 10^2$ at an overdensity
$\delta\equiv\rho/\bar{\rho} \sim 10$ (or $n_{\rm H} \sim
10^{-4}~\cm^{-3}$ at $z=3$) up to $\ga 10^3$ at $\delta \sim 1$.  We
find that this renders ionization corrections as used in Paper II
unreliable for \SiIV, because very small errors in the inferred \HI\
optical depth lead to large errors in the recovered silicon
abundance. Rather than applying ionization corrections to the data, we
therefore concentrate on directly comparing the predictions of our
hydrodynamical simulation to the observed spectra. Using the
measurements of the carbon distribution obtained in Paper II, we find
the value of [Si/C] that best fits the data.

\begin{figure*}
\epsscale{1.0} 
\plottwo{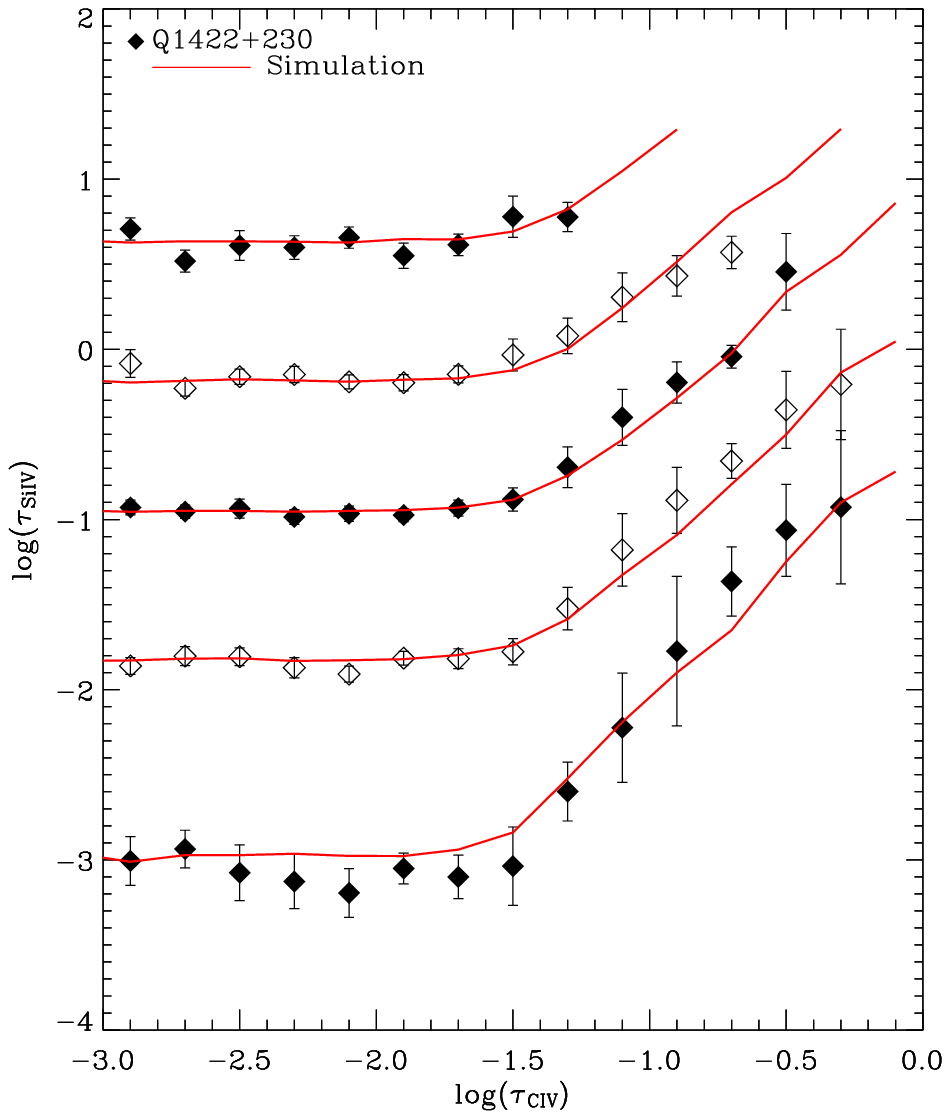}{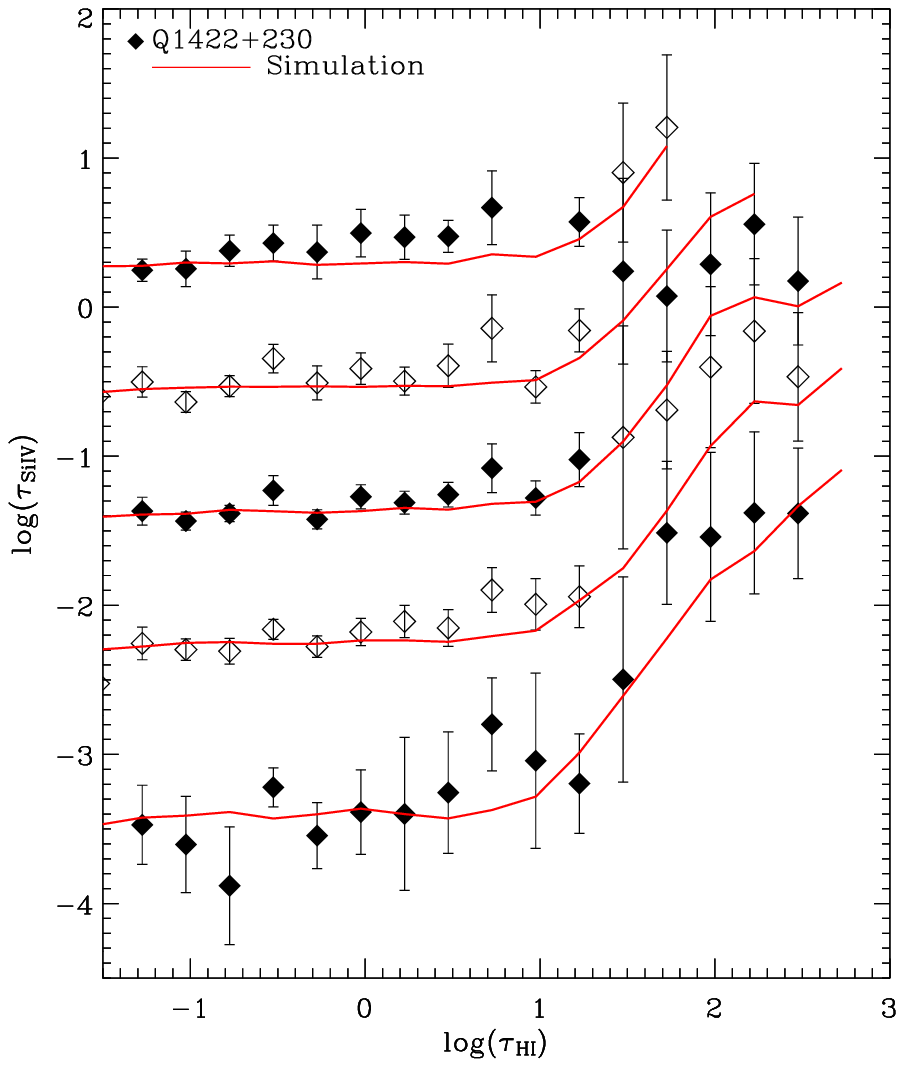}
\figcaption[]{SiIV optical depths in bins of $\tciv$ (left panel) and
$\thi$ (right panel) for Q1422+230.  From bottom to top, the points
are the median, 69th, 84th, 93rd and 97th percentiles. Solid lines represent
predictions from simulations with $\langle {\rm [C/H]}\rangle = -3.8 +
0.65 \delta$, $\sigma=0.70$, and [Si/C]$=\log 5.$ For clarity, both
observed points and simulated lines are plotted with vertical
offsets of (from bottom to top) 0.0, 0.5, 1.0, 1.5, and 2.0 dex. Note
that for the right panel the spectra have been smoothed by $7.5\,\kms$ (see
\S~\ref{sec-sivhi}).
\label{fig:sivscat}}
\end{figure*}

For each observed quasar we generate 50 corresponding simulated
spectra with the same noise properties, wavelength coverage,
pixelization, instrumental broadening, and excluded regions. We apply
to the simulated spectra the same automated continuum fitting routine
as was used for the observations (Paper II, \S~5.1, step 1; see also
Paper I, \S~4.1) to reduce any differences in the continuum fitting
errors between simulated and observed spectra.

Before generating the spectra, we first impose a carbon distribution
in the simulation that is consistent with the measurements presented
in Paper II. In that paper we found that for a given overdensity and
redshift, the probability distribution for the carbon abundance is
well-described by a lognormal function, and we presented fits of the
parameters of the distribution (i.e., the median and the width) as a
function of overdensity and redshift for four different UVBs (Paper
II, Eqn.\ 8 and Table 2). For the QG and Q ionizing backgrounds these
fits are consistent with no evolution and we use the values for $z=3$;
for models QGS and QGS3.2 we use, for each quasar, the median redshift
of the analyzed region. We divide each simulation snapshot into $10^3$
cubes, and assign the gas particles in each section a carbon abundance
of $$[{\rm C}/{\rm H}]=\alpha + \beta(z-3)+\gamma(\log\delta-0.5)+s,$$
where $s$, which is the same for all particles in the sub-volume, is
drawn at random from a lognormal distribution with mean 0 and variance
$\sigma([{\rm C}/{\rm H}]) = 0.70$, and $\delta$ is the overdensity of the
particle. For QG, $\alpha=-3.47$, $\gamma=0.65$, and $\beta=0$; for Q,
$\alpha=-2.91$, $\gamma=0.17$, and $\beta=0$ (see Paper II, Table~2
for values used for other UVBs).  Silicon abundances are assigned by
assuming a {\em constant} value of [Si/C].  (The assumption that
silicon tracks carbon perfectly is tested below.)

For each observed and simulated quasar spectrum, optical depths for
pixels corresponding to absorption by \HI, \CIV, \SiIV\ and \SiIII\
are extracted as described in~\S\ref{sec-recovery}. We may then plot
various percentiles (such as the median) of the $\tau_{\rm SiIV}$
distribution for each bin in $\tciv$.  The same procedure can be
applied to correlate \SiIV\ with \HI, or \SiIII\ with \SiIV. For each
\CIV\ (or other) bin, errors on the observed points are calculated by
bootstrap resampling the spectrum, i.e., we divide the Ly$\alpha$
forest region of the spectrum into chunks of 5~\AA\ which are bootstrap
resampled to form a new realization of the spectrum (see Paper II
\S5.1, step 3).  We require that in each bin there be at least five
pixels above the percentile being computed, and that at least 25
pixels and five chunks contribute to the bin so that the errors are
reliable; otherwise the point is discarded.  For the simulations we
compute 50 synthetic spectra and each percentile in each bin is set
equal to the median of the 50 realizations, with errors given by
bootstrap resampling the 50 simulated spectra. We require each bin of
each realization to have at least five pixels above the given
percentile, and to have at least five pixels and one chunk contributing to the
bin, and we discard medians computed with fewer than five acceptable
realizations.

For each percentile, the correlation disappears below some \CIV\
optical depth $\tau_c$ at a value $\tau_{\rm min}$ that is determined
by noise, continuum fitting errors, and contamination by other lines.
These may be corrected for by subtracting $\tau_{\rm min}$ from the
binned optical depths, thus converting points at $\thi < \tau_c$ into
upper limits (Paper II, section 5.1, step 4, and Fig.\ 4). For each
realization, we compute $\tau_{\rm min}$ as the given percentile of
the optical depth for pixels with $\thi < \tau_c$, where $\tau_c=0.01$
is chosen as the \CIV\ or \SiIV\ optical depth below which we never
see a correlation.\footnote{In Paper II we used functional fits to
determine $\tau_c$. For SiIV the correlations are generally less
strong than for CIV and we fix $\tau_c$ ``by hand''. For bins in
$\thi$, we use $\tau_c=1$.}  The error on $\tau_{\rm min}$ for each
realization is computed by bootstrap-resampling the quasar
spectrum. When the realizations are combined, the value of $\tau_{\rm
min}$ is computed as the median among the realizations, and the error
on this value is computed by bootstrap-resampling the realizations.

The binned optical depth percentiles are then compared directly to
the simulations, point by point.  Because of slight differences in
contamination, accuracy of continuum fitting, and noise, $\tau_{\rm
min}$ for a given QSO may differ slightly from that of the
simulations.  Because we do not want to compare noise with noise, we
either: A) subtract $\tau_{\rm min}$ from both simulations and
observations before comparing,\footnote{As in Paper II, we propagate
the asymmetric errors for points in which $\tau_{\rm min}$ has been
subtracted by computing a fine grid of errors (e.g., $\pm 0.01\sigma,
0.02\sigma, \ldots$).} or B) add a constant to the simulated optical
depths, chosen to minimize the difference between the simulations and
observations for $\thi < \tau_c$; in this case we compare only points
at $\thi \ge \tau_c$

\section{Results}
\label{sec-resrel} 

Before presenting the results from our full sample, we
  illustrate the optical depth statistics by showing some results from
  our QSO with the strongest signal, Q1422+230.

\subsection{Results for Q1422+230}

Fig.~\ref{fig:sivscat} shows several percentiles of $\tau_{\rm SiIV}$,
binned according to their corresponding optical depth in \CIV\ (left)
and \HI\ (right) for Q1422+230.  The bottom points are medians, and
the next four sets (with vertical offsets of 0.5,1.0, 1.5, and 2.0
dex) represent percentiles 69.146, 84.134, 93.319, and 97.725.  These
correspond to x=0, 0.5,1,1.5 and 2$\sigma$ values of the cumulative
gaussian probability function $f(x) = {1 \over \sqrt{2\pi}}
\int_{-\infty}^{x} e^{-t^2/2\sigma^2} dt$ (so that, for example, a
distribution of $\tsiiv$ that is lognormal with width $1\,$dex would
give $1\,$ dex higher metallicity in the 84.134th percentile than in the
median).  The solid lines are predictions from the simulations with
[Si/C]$=\log 5=0.70$ uniformly, median C metallicity $\langle {\rm
[C/H]}\rangle = -3.8 + 0.65 \delta$ and a width of $\sigma([{\rm
C/H}])=0.70$\,dex in the (lognormal) metallicity
distribution.\footnote{These values are taken from the overall surface
fits from Paper II; $\langle {\rm [C/H]}\rangle$ is evaluated at $z=3$
and the small redshift evolution is neglected; $\sigma([{\rm C/H}])$
is evaluated at $z=3$ and $\log\delta=0.75$, with both $\delta$- and
$z$-dependences neglected.}

The left panel shows that the simulations with the QG UVB and
[Si/C]$=\log 5$ match the observations very well.  This can be
quantified by computing, for all points with $\log\tciv \ge
\log\tau_c=-2.0$, the $\chi^2$ difference between the simulated and
observed data points.\footnote{We take into account both simulated and
observed errors, and adjust $\tau_{\rm min}$ of the simulations as
discussed in~\S~\ref{sec-compmeth}.} We obtain $\chi^2/{\rm
d.o.f.}=5.5/9,4.0/9,2.2/8,$ and 7.4/7 for the median, 0.5$\sigma$,
1$\sigma$ and 1.5$\sigma$ percentiles, respectively.\footnote{The
rather low reduced $\chi^2$ values may indicate some correlation
between neighboring points, but they largely result from a slight
overestimate of the errors near $\tau_c$; see Paper II, \S~7.}  The
good fit of the median indicates that a uniform [Si/C] $\sim \log 5$
model fits well (though slightly higher [Si/C] are favored;
see below), and that the data do not require any strong trend of
[Si/C] with density (for which $\tciv$ is a proxy).  Comparison of the
higher percentiles can help constrain scatter in $\tsiiv/\tciv$, which
depends on the scatter in both [Si/C] and the ionization
correction, and it is clear that that simulations with uniform [Si/C]
and a uniform UVB fit the observations well; this is
quantified below using our full data sample.

Comparison between simulated and observed $\tsiiv$ versus $\thi$
(right) is somewhat less straightforward, as it depends both on [Si/C]
and on the assumed distribution of carbon.  Encouragingly, we find
that for Q1422+230 a model with uniform [Si/C] (and carbon
distribution as derived in Paper II from the full sample) fits the
observed medians fairly well.  The predicted higher percentiles are,
however, a bit low; this is partly because the Q1422+230 \CIV\
absorbers happen to exhibit slightly more metallicity scatter
($\approx 0.81\,$dex; see Paper II, \S~5.2) than the level derived
from the full sample.

In summary, the observed Q1422+230 \SiIV\ optical depths, and their
comparison to simulations show that: 1. A uniform [Si/C]$\sim \log 5$
is favored, 2. There is no evidence for a trend of [Si/C] with
density, 
3. There is no
evidence for scatter in $\tsiiv/\tciv$ beyond that included in the
simulation with uniform [Si/C] and a uniform UVB.

\subsection{$\tsiiv$ versus $\tciv$ for the full sample}

\begin{figure}
\epsscale{1.00} \plotone{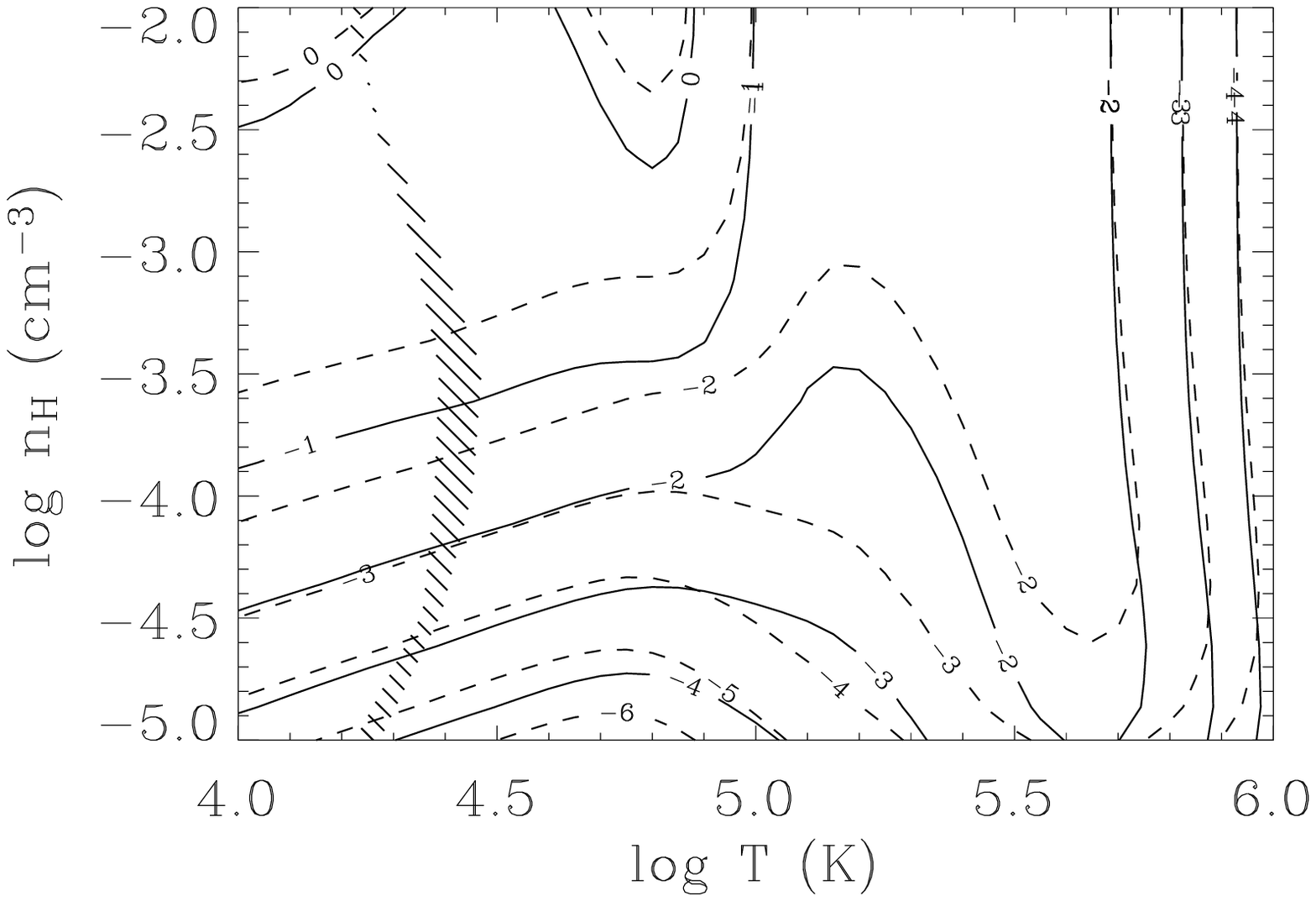} \figcaption[]{Predicted optical depth
ratios $\log \tsiiv/\tciv$ as a function of the temperature and the
hydrogen number density, assuming a solar ratio of Si/C. Solid
(dashed) contours are for the UV-background model QG (Q) at $z=3$. The
hatched region indicates the temperature range containing 90\% of the
particles at the given density.
\label{fig:sivcivpred}}
\end{figure}

\begin{figure*}
\plotone{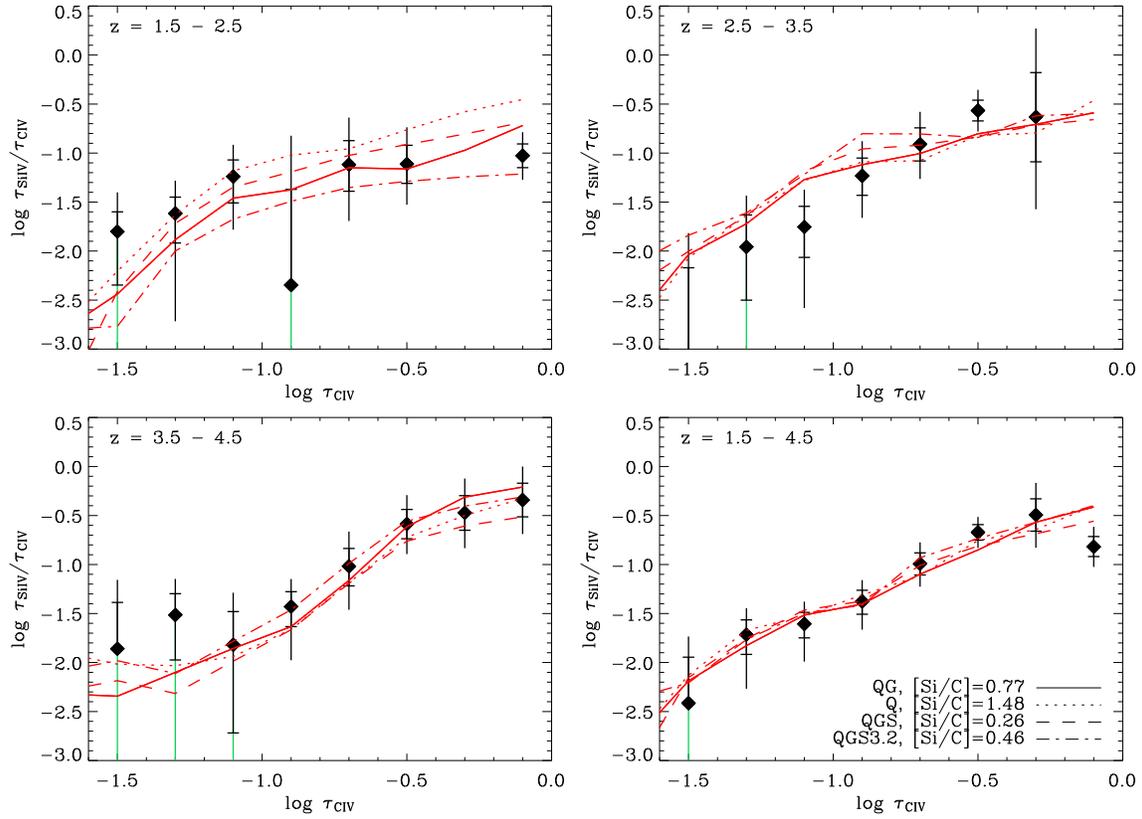} \figcaption[]{Rebinned median $\log(\tsiiv/\tciv)$
vs. $\log\tciv$ in cuts of $z$ for the combined QSO sample. The first
three panels show bins centered at $z=2.0,3.0$ and 4.0 with width
$\Delta z = 1.$; the bottom-right panel shows combined data from all
redshifts.  Lines represent corresponding rebinned simulation points
(with errors suppressed, and with [Si/C] chosen to minimize the
$\chi^2$) using different UVB models.
\label{fig:sc_zcuts}}
\end{figure*}

\begin{figure*}
\plotone{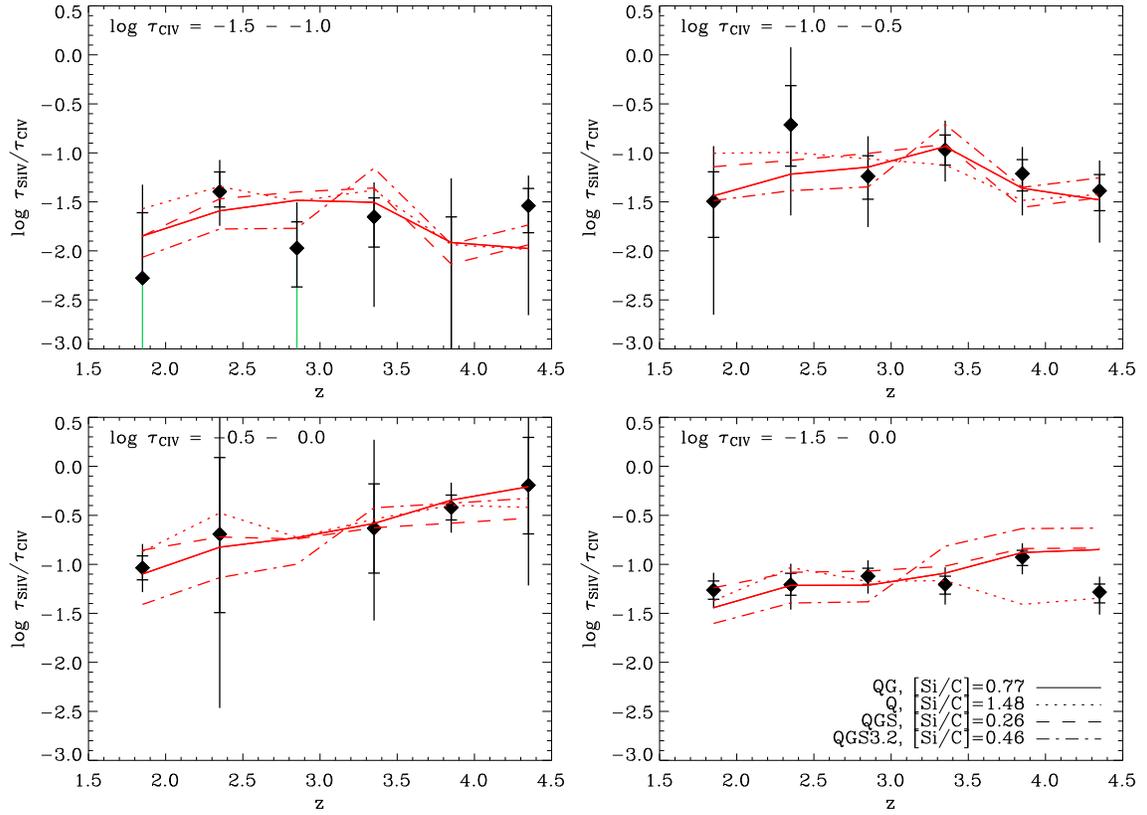}
\figcaption[]{Rebinned median $\log(\tsiiv/\tciv)$ vs. $z$ in cuts of $\tciv$
for the combined QSO sample. The first three panels show bins centered
at $\log\tciv=-1.25,-0.75$ and -0.25 with width $0.5$\,dex; the
bottom-right panel shows data for all $\tciv$ combined.  Lines represent
corresponding rebinned simulation points (with errors suppressed, and
with [Si/C] chosen to minimize the $\chi^2$) using different UVB models.
\label{fig:sc_dcuts}}
\end{figure*}

\begin{deluxetable*}{lcccccccccccccc}
\tabletypesize{\tiny}
\tablewidth{0pc}
\tablecaption{Recovered SiIV optical depths
\label{tbl:recod}}
\tablecomments{Columns 1 and 2 contain the quasar name and the
recovered CIV optical depth respectively. Columns 3 and 4 contain the
31st percentile of the recovered SiIV optical depth and the $1\sigma$
error on this value. The remaining columns show the same quantities
for higher percentiles (50th, 69th, 84th, 93th, and 98th). A CIV
optical depth of $\log\tau_{\rm CIV}=-9$ indicates that the
corresponding $\tau_{\rm SiIV}$ are the $\tau_{\rm min}$ values.  The
complete version of this table, including all quasars, is in the
electronic edition of the Journal. The printed edition contains only
data for Q1422+230.}
\tablehead{
\colhead{QSO} & \colhead{$\log\tau_{{\rm CIV}}$} &
\multicolumn{2}{c}{$\log\tau_{{\rm SiIV}}, -0.5\sigma$} &
\multicolumn{2}{c}{$\log\tau_{\rm SiIV}, {\rm median}$} &
\multicolumn{2}{c}{$\log\tau_{\rm SiIV}, +0.5\sigma$} &
\multicolumn{2}{c}{$\log\tau_{\rm SiIV}, +1\sigma$}&
\multicolumn{2}{c}{$\log\tau_{\rm SiIV}, +1.5\sigma$}&
\multicolumn{2}{c}{$\log\tau_{\rm SiIV}, +2\sigma$}}
\startdata
Q1422+230 & -9.000 & \nodata & \nodata  & -3.047 &  0.088   & -2.343 &  0.035   & -1.959 &  0.031   & -1.675 &  0.035   & -1.354 &  0.047  \\
Q1422+230 & -3.900 & -3.371 &  0.579   & -2.530 &  0.153   & -2.115 &  0.085   & -1.898 &  0.071   & \nodata & \nodata  & \nodata & \nodata \\
Q1422+230 & -3.700 & \nodata & \nodata  & -2.827 &  0.197   & -2.183 &  0.092   & -1.817 &  0.111   & -1.622 &  0.105   & \nodata & \nodata \\
Q1422+230 & -3.500 & \nodata & \nodata  & -2.847 &  0.199   & -2.308 &  0.076   & -1.978 &  0.074   & -1.765 &  0.049   & \nodata & \nodata \\
Q1422+230 & -3.300 & \nodata & \nodata  & -3.214 &  0.146   & -2.492 &  0.114   & -2.054 &  0.071   & -1.786 &  0.127   & -1.310 &  0.141  \\
Q1422+230 & -3.100 & \nodata & \nodata  & -2.879 &  0.139   & -2.270 &  0.066   & -1.918 &  0.058   & -1.605 &  0.070   & -1.326 &  0.072  \\
Q1422+230 & -2.900 & \nodata & \nodata  & -3.007 &  0.143   & -2.361 &  0.049   & -1.929 &  0.044   & -1.583 &  0.082   & -1.293 &  0.065  \\
Q1422+230 & -2.700 & \nodata & \nodata  & -2.936 &  0.110   & -2.303 &  0.056   & -1.954 &  0.039   & -1.729 &  0.046   & -1.482 &  0.064  \\
Q1422+230 & -2.500 & \nodata & \nodata  & -3.076 &  0.164   & -2.306 &  0.052   & -1.936 &  0.055   & -1.661 &  0.044   & -1.390 &  0.088  \\
Q1422+230 & -2.300 & \nodata & \nodata  & -3.127 &  0.160   & -2.371 &  0.059   & -1.985 &  0.043   & -1.648 &  0.048   & -1.402 &  0.069  \\
Q1422+230 & -2.100 & \nodata & \nodata  & -3.194 &  0.144   & -2.409 &  0.047   & -1.966 &  0.045   & -1.691 &  0.041   & -1.344 &  0.062  \\
Q1422+230 & -1.900 & \nodata & \nodata  & -3.050 &  0.090   & -2.315 &  0.040   & -1.974 &  0.039   & -1.697 &  0.048   & -1.450 &  0.075  \\
Q1422+230 & -1.700 & \nodata & \nodata  & -3.100 &  0.128   & -2.318 &  0.059   & -1.933 &  0.047   & -1.647 &  0.050   & -1.386 &  0.063  \\
Q1422+230 & -1.500 & \nodata & \nodata  & -3.037 &  0.230   & -2.277 &  0.077   & -1.882 &  0.068   & -1.533 &  0.094   & -1.221 &  0.121  \\
Q1422+230 & -1.300 & \nodata & \nodata  & -2.599 &  0.173   & -2.024 &  0.125   & -1.693 &  0.120   & -1.421 &  0.104   & -1.222 &  0.086  \\
Q1422+230 & -1.100 & -3.610 &  0.606   & -2.223 &  0.321   & -1.678 &  0.213   & -1.400 &  0.165   & -1.194 &  0.143   & \nodata & \nodata \\
Q1422+230 & -0.900 & -3.351 &  0.772   & -1.773 &  0.439   & -1.388 &  0.193   & -1.195 &  0.121   & -1.069 &  0.118   & \nodata & \nodata \\
Q1422+230 & -0.700 & -1.873 &  0.313   & -1.364 &  0.203   & -1.156 &  0.102   & -1.044 &  0.067   & -0.930 &  0.095   & \nodata & \nodata \\
Q1422+230 & -0.500 & -1.315 &  0.497   & -1.063 &  0.270   & -0.856 &  0.226   & -0.545 &  0.225   & \nodata & \nodata  & \nodata & \nodata \\
Q1422+230 & -0.300 & -1.264 &  0.813   & -0.928 &  0.450   & -0.706 &  0.325   & \nodata & \nodata  & \nodata & \nodata  & \nodata & \nodata \\
Q1422+230 & -0.100 & \nodata & \nodata  & \nodata & \nodata  & \nodata & \nodata  & \nodata & \nodata  & \nodata & \nodata  & \nodata & \nodata \\
\enddata
\end{deluxetable*}

To place more quantitative constraints on [Si/C], test for evolution,
and make a more detailed comparison with several UVB models, we have
combined the data points obtained from our entire sample.(These points
are tabulated in Table 1.)  Figure~\ref{fig:sc_zcuts} shows $\log
\tsiiv/\tciv$ versus\ $\log \tciv$, in bins of $z$. To generate these
points, we begin with $\tsiiv$ values binned in $\tciv$, as in
Fig.~\ref{fig:sivscat}.  We then subtract from each the ``flat level''
$\tau_{\rm min}$ for that QSO to adjust for noise, contamination etc.\
(see \S~\ref{sec-meth}), then divide by the central value of the
$\tciv$ bin.  These points, gathered from all QSOs, are rebinned by
$\chi^2$-fitting, for each $\tciv$ range indicated in
Fig.~\ref{fig:sc_zcuts}, a constant level to all of the points in the
specified redshift range. The errors represent 1- and 2-$\sigma$
confidence intervals ($\Delta\chi^2=2$ and $\Delta\chi^2=4$).

The plotted lines indicate predictions from the simulations for our
different UVBs; our fiducial model, QG, is shown in solid lines.  For
each background, we generate simulated $\tsiiv/\tciv$ points in the
same way as we did for the observations, but averaging over 50
simulated realizations as described in~\S~\ref{sec-compmeth}.  We then
calculate a $\chi^2$ between all valid observed original (not
rebinned) points and the corresponding simulated points.\footnote{Even
using 50 realizations, it may occasionally happen that a simulated
bins fails to have enough pixels for at least five realizations, and
so is undefined; in this case the observed point is discarded as
well.}  Because we use 50 simulated realizations, the simulation
errors are almost always negligible compared to the observed errors,
but they are still taken into account by calculating the total $\chi^2$
using the formula:
\begin{equation}
\chi^2=\sum_i \left[\left({X_{\rm obs}-X_{\rm sim}\over\sigma_{\rm
      obs}}\right)^{-2}+\left({X_{\rm obs}-X_{\rm sim}\over\sigma_{\rm
      sim}}\right)^{-2}\right]^{-1},
\label{eq:chi2}
\end{equation}
where $X\equiv\tsiiv/\tciv$ and $\sigma$ is the error in this
quantity.  We then add a constant offset to the simulated points
(which corresponds to scaling [Si/C]) such that $\chi^2$ is
minimized.  In each panel the lines connect the scaled, rebinned
simulation points.

The first clear result is that there is a strong trend of
$\log(\tsiiv/\tciv)$ with $\log\tciv$, from $\la -2$ at $\log\tciv\sim
-1.5$ to $\ga -1.0$ at $\log\tciv \ga -0.5$.  This previously unnoted
correlation is expected if [Si/C] is constant and $\tciv$ correlates
with density, since (as shown in Fig.~\ref{fig:sivcivpred})
$\tsiiv/\tciv$ increases rapidly with gas density.  As shown by the
simulation lines, this trend is reproduced in all redshift bins by the
simulations for all of our UVB models.  Comparing the first three
panels shows that while $\tsiiv/\tciv$ correlates strongly with
$\tciv$, there is little dependence on redshift.  This can be seen
more clearly in the first three panels of Fig.~\ref{fig:sc_dcuts},
which show $\log(\tsiiv/\tciv)$ versus $z$ in bins of $\tciv$. There is
no evidence, in either the simulated or the observed points, for evolution
in $\tsiiv/\tciv$, except perhaps for a slight increase in
$\tsiiv/\tciv$ with increasing $z$ at the highest densities.

\begin{deluxetable}{lcc} 
\tabletypesize{\scriptsize}
\tablewidth{0pc} \tablecaption{Best fit [Si/C] and $\chi^2/{\rm d.o.f.}$
\label{tbl:fits}}  
\tablehead{
\colhead{UVB model} & 
\colhead{best fit [Si/C]} &
\colhead{$\chi^2$/d.o.f.}}
\startdata
QG & $0.77^{+0.05}_{-0.05}$ & 65.7/115 \\
Q & $1.48^{+0.05}_{-0.06}$ & 65.6/115 \\
QGS & $0.26^{+0.06}_{-0.07}$ & 73.8/115 \\
QGS3.2 & $0.46^{+0.10}_{-0.08}$ & 85.2/115 \\ 
\enddata

\end{deluxetable} 

\begin{figure*}
\plotone{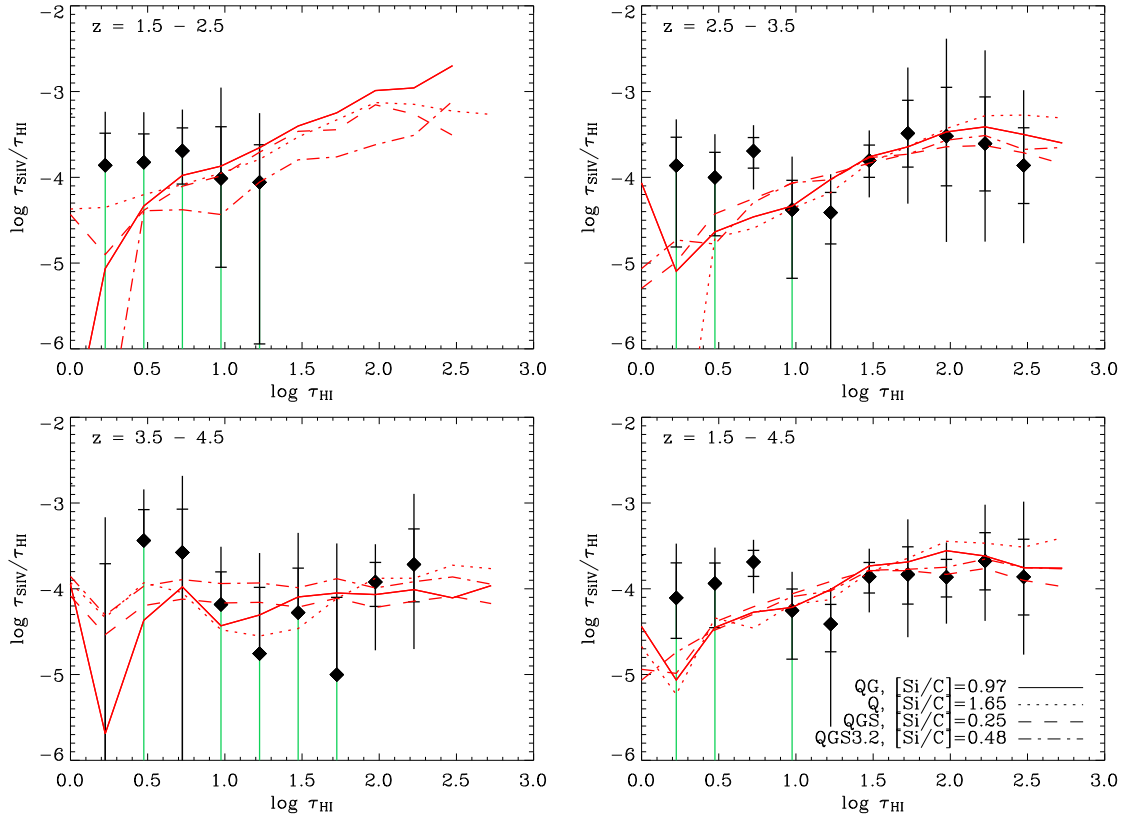} \figcaption[]{Median $\log(\tsiiv/\thi)$ vs.\
$\log\thi$ in bins of $z$ for the combined QSO sample. The first three
panels show bins centered at $z=2,3,$ and 4 with width $\Delta z=1$;
the bottom-right panel shows combined data for all redshifts.  The
lines represent corresponding simulation points (with errors
suppressed) using different UVB models.
\label{fig:sh_zcuts}}
\end{figure*}

The observed trends in $\log(\tsiiv/\tciv)$ are reproduced well by the
simulations. Because $\tsiiv/\tciv$ scales with Si/C, the offset in
$\tsiiv/\tciv$ obtained by minimizing the 
$\chi^2$ (Eq.~\ref{eq:chi2}) against the observations can be used to
compute the best fit [Si/C].  For our fiducial UVB model QG, the
simulated spectra were generated with [Si/C]=0.70, we find an offset
of $+0.07$\,dex (implying a best-fit [Si/C]=0.77) with $\chi^2/{\rm
d.o.f.}$=65.7/115.  As we found in Paper II and for Q1422+230 above,
the reduced $\chi^2$ is somewhat low; in Paper II we showed that his
was largely due to a slight overestimate of the errors at low-$\tciv$.

The fitted [Si/C] values and corresponding $\chi^2/{\rm d.o.f.}$, are
listed in Table~\ref{tbl:fits}, with errors
computed by bootstrap-resampling the quasars used in the $\chi^2$
minimization.  For our fiducial model, QG, the best fit
[Si/C]$=0.77^{+0.05}_{-0.05}$.  The quasar-only background Q (which is
probably too hard; see Paper II) gives a higher values of
[Si/C]$=1.48^{+0.05}_{-0.06}$, and the softer QGS and QGS3.2
backgrounds give lower values than QG by $\approx 0.3-0.5$\,dex. Note
that the QGS background is unrealistically soft at $z \lsim 3$ (see
Paper II), but the QGS3.2 background may be plausible.

Because the $\tsiiv/\tciv$ ratio depends on both [Si/C] and the shape
of the UVB, its measurement can be used to study the evolution of the
UVB under the assumption that [Si/C] is constant. \citet[ see also
Songaila \& Cowie 1996]{1998AJ....115.2184S} has measured the median
\SiIV/\CIV\ ratio versus $z$ for \CIV\ systems of column density
$5\times 10^{12}\,{\rm cm^{-2}} \le N({\rm CIV}) \le 10^{14}\,{\rm
cm^{-2}}$ and found evidence for strong evolution, as well as a sharp
break in $\tsiiv/\tciv$ at $z = 3.0.$ This was interpreted as evidence
for a sudden softening of the UVB at $z > 3$.  However, in a recent
study~\citet{boksen}, in agreement with their earlier work and that
of~\citet{2002A&A...383..747K} find {\em no} evolution in
$\tsiiv/\tciv$ in their sample of $10^{12}\,{\rm cm^{-2}} \le N({\rm
CIV}) \la 3\times 10^{14}\,{\rm cm^{-2}}$ absorbers.  As discussed
above (see Fig.~\ref{fig:sc_dcuts}), we see no evidence for evolution
in the median $\tsiiv/\tciv$ stronger than that predicted by the
simulations, in any of our $\tciv$ bins, or when all $\tciv$ values
are combined (Fig.~\ref{fig:sc_dcuts}, bottom-right).

Because $\tsiiv/\tciv$ varies by $\sim 1.5\,$dex in correlation with
$\tciv$, its evolution is best assessed by using only a small window
in $\tciv$; otherwise evolution in the weight provided by each $\tciv$
-- whether due to selection effects or evolution in the distribution
of $\tciv$ -- may lead to apparent evolution in $\tsiiv/\tciv$.  In
our analysis, in which we use small cuts in $\tciv$,\footnote{In the
bottom right panel of Fig.~\ref{fig:sc_dcuts}, we combine all $\tciv$
values.  Although the observations and simulations should have similar
weightings by \CIV, some differences may remain and the comparison is
less reliable than if cuts in $\tciv$ are made.} we see no evidence
for evolution in the UVB beyond that in the smoothly changing QG and Q
models; indeed, the QGS3.2 model with an artificial change in softness
at $z=3.2$ is disfavored by our data, having a significantly higher
$\chi^2$ than either the QG or Q models.

The simulations we employ assume a constant and uniform [Si/C]; but
because the formation mechanism for Si and C may be different this
need not be the case.  We can, however, observationally constrain the
scatter in [Si/C] by repeating our determination of it using different
percentiles in $\tsiiv/\tciv$.  If the probability distribution of
[Si/C] for fixed $\tciv$ is, like that of [C/H] and [Si/H] for fixed
$\thi$, lognormal, then we can directly constrain the width
$\sigma({\rm [Si/C]})$ of the distribution by comparing [Si/C] derived
for different percentiles.  For the 69th, 84th, 93rd and 97th
percentiles, we obtain [Si/C]=$0.73^{+0.05}_{-0.06}$,
$0.64^{+0.07}_{-0.07}$, $0.58^{+0.13}_{-0.04}$, and
$0.78^{+0.10}_{-0.14}$, respectively. This translates into a rough
2$\sigma$ upper limit of $\sigma({\rm [Si/C]}) \la
0.04\,$dex.\footnote{This is only a rough error estimate as it assumes
that the points are fully independent, which does not hold because the
points correspond to a ranking.}  As a rough test of this upper
limit, we have generated simulated spectra with a {\em median} ${\rm
[Si/C]=0.77}$ and the usual C distribution, but with a lognormal
scatter in [Si/C] of width $\sigma({\rm [Si/C]})$.  If we add the
$\chi^2$ for the 69th, 84th and 93rd percentiles for quasars at $z \ge
3$ (there is little useful information on the upper percentiles from
$z < 3$), we find $\chi^2/{\rm d.o.f.}=198.0/185$ for no scatter in
[Si/C], and $\chi^2=198.2, 203.4, $ and 223.0 for $\sigma({\rm
[Si/C]})=0.1,0.2, $and 0.4 dex, respectively.  Because the percentiles
are correlated these cannot be correctly translated directly into
confidence limits; however they suggest that the data are compatible with
$\sigma({\rm [Si/C]})=0.1$, but probably not with $\sigma({\rm
[Si/C]})=0.2$ and almost certainly not with $\sigma({\rm
[Si/C]})=0.4$.  Thus, contrary to the large scatter in [C/H],
$\sigma({\rm [C/H]})\approx 0.7$\,dex found in Paper II, there
appears to be very little scatter in [Si/C].

We may also subdivide our sample by redshift and density to test the
dependence of [Si/C] on these.  First, computing [Si/C] using only
spectra that have a median absorption redshift ${\rm med}(z) > 3.0$
(see Paper II, Table 1) yields [Si/C]=$0.76^{+0.05}_{-0.07}$, versus
[Si/C]=$0.79^{+0.10}_{-0.08}$ using the spectra with ${\rm med}(z) <
3.0$.  The [Si/C] values inferred from the redshift subsamples are
also (marginally) consistent for the Q and QGS UVBs, but
{\em in}consistent for QGS3.2; the latter would imply a jump 
from [Si/C]$=0.77^{+0.05}_{-0.09}$ at $z > 3$ to
[Si/C]$=0.32^{+0.08}_{-0.13}$ at $z < 3$.  This is a second way of
seeing that our data disfavors a sudden change in UVB hardness near
$z=3$, assuming that [Si/C] is constant.

To test for variation in [Si/C] with overdensity $\delta$ we have
recomputed [Si/C] using only pixels with $\thi$ corresponding to
$\delta < 20$ or $\delta > 20$ (using the $\thi$-$\delta$ conversion
of Paper II, Fig. 2).  We obtain [Si/C]=$0.87^{+0.19}_{-0.10}$ for the
low-density sample, versus [Si/C]=$0.73^{+0.03}_{-0.05}$ at high
density.  This difference could imply either that [Si/C] is higher at
low densities, or that the UVB is softer than we have assumed (see
Fig.~\ref{fig:sivcivpred}); but the effect is only significant at the
$\approx 1.3\sigma$ level.

\subsection{$\tsiiv$ versus  $\thi$ for the full sample}
\label{sec-sivhi}

While the $\tsiiv/\tciv$ ratios give the most direct constraints on
[Si/C], it is also useful to examine $\tsiiv/\thi$: comparing the
simulated to the observed $\tsiiv/\thi$ ratios gives, in principle, a
second estimate of [Si/C].  This inference, however, is less reliable
than that from $\tsiiv/\tciv$ for two reasons.  First, it includes the
uncertainty in [C/H], which is largest at the relatively high
densities at which (because of the small \SiIV/Si fraction at lower
densities) \SiIV\ is best-detected.  Second, Si absorption lines have
a much smaller thermal width than hydrogen lines, and are only
observable in relatively high-density gas (where Hubble broadening is
small).  This leads to significant differences between the
\SiIV-weighted and \HI-weighted densities (see Paper II, \S~4.3, Paper
I, \S~5.3), i.e., the \SiIV\ and \HI\ absorption do note arise from
exactly the same gas.  The effect of this is to skew (and weaken) the
correlation of $\tsiiv$ with $\thi$ in a way that depends on the
\HI\ column-density distribution. Unfortunately, as discussed
by~\citet{2002ApJ...578L...5T}, the simulation does not exactly
reproduce the observed $N({\rm HI})$ distribution at the high-column
density end (which is most important for the present study), so the
``skewing'' may be somewhat different in observed and simulated
correlations, and render conclusions about [Si/H] unreliable.

The problems arising from differential line broadening can be
partially remedied by smoothing the spectra so that the minimal line
widths of all species become similar; experimentation shows that smoothing
the spectra by convolving them with a Gaussian with FWHM of $\approx
5-10\,\kms$ significantly increases the strength of the signal, and a
smoothing of 7.5\,$\kms$ has been adopted for the calculations shown
in Figs.~\ref{fig:sivscat} and~\ref{fig:sh_zcuts}.
Fig.~\ref{fig:sh_zcuts} shows $\log\tsiiv/\thi$ versus $\log\thi$ in
bins of $z$ for our combined sample. Lines again connect the
corresponding simulation points (with an overall scaling to best match
the observations) that reproduce the observed trends in $z$ and
$\thi$. The scalings correspond to best-fit [Si/C] values of
$0.97\pm0.08$, $1.65\pm0.08$, $0.25\pm0.07$ and $0.48\pm0.09$ for QG,
Q, QGS, and QGS3.2 respectively, slightly higher than the values found
above using $\tsiiv/\tciv$. However, as explained above, these values
depend on the degree of smoothing: for example, with no smoothing, we
recover values $\approx 0.1-0.3\,$dex higher yet.  Since it is unclear
which smoothing level gives the correct results, the inferences of
[Si/C] using $\tsiiv/\thi$ should not be relied on. 

The same difference in thermal width affects inferences using
\SiIV/\CIV, but at a lower level; whether or not we smooth by
7.5$\kms$ changes the inferred [Si/C] by $\sim 0.1$\,dex, which is a
reasonable estimate of the induced systematic error.

\begin{figure}
\epsscale{1.00} \plotone{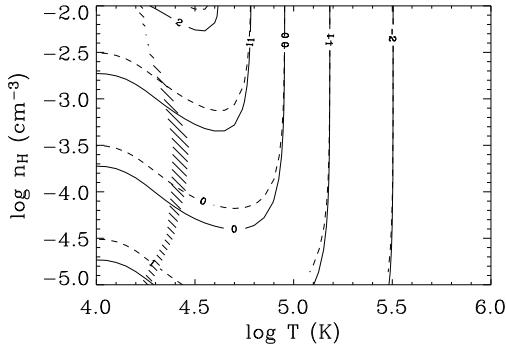} \figcaption[]{Predicted optical depth
ratios $\log \tau_{\rm SiIII}/\tsiiv$ as a function of the temperature
and the hydrogen number density. Solid (dashed) contours are for the
UV-background model QG (Q) at $z=3$. The hatched region indicates the
temperature range containing 90\% of the particles at the given
density.
\label{fig:siiisivpred}}
\end{figure}

\subsection{$\tsiiii$ versus  $\tsiiv$ for the full sample}

A final correlation that we have examined is that of $\tsiiii/\tsiiv$
with $\tsiiv$.  Fig.~\ref{fig:siiisivpred} shows the predicted
$\tsiiii/\tsiiv$ versus the temperature $T$ and density $n_{\rm H}$ of
the absorbing gas.  For high $T$, $\tsiiii/\tsiiv$ becomes
independent of $n_{\rm H}$, and declines rapidly with $T$.  Thus, the
presence of \SiIII\ can be used to constrain the temperature of the
gas providing \SiIV\ absorption.  In Paper II a similar test was
carried out using the $\tau_{\rm CIII}/\tciv$ ratio, yielding the
constraint $T < 10^{5.0}\,$K.

\begin{figure}
\epsscale{1.0} 
\plotone{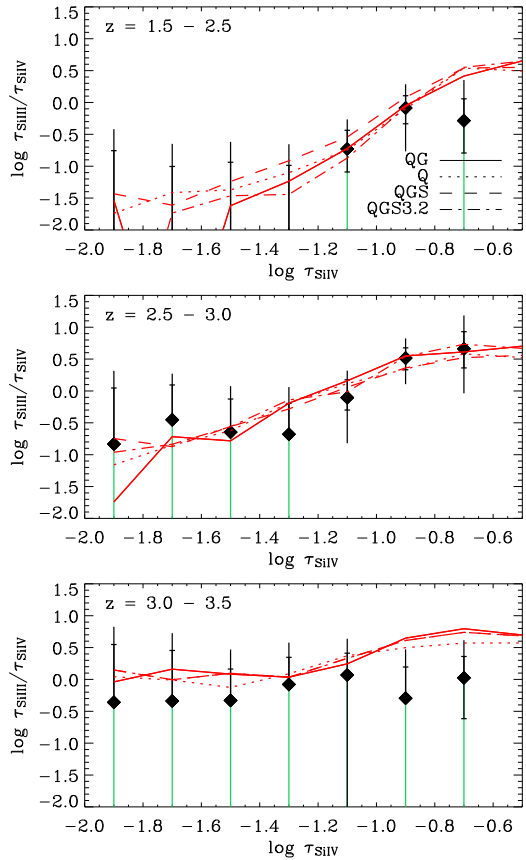} \figcaption[]{Ratios of median
$\tau_{\rm SiIII}/\tau_{\rm SiV}$ versus $\tsiiv$ in bins of $z$ for the
full sample.  The lines represent corresponding simulation points
(with errors suppressed) using different UVB models.
\label{fig:sisi_zcuts}}
\end{figure}

Fig.~\ref{fig:sisi_zcuts} shows $\tsiiii/\tsiiv$ versus $\tsiiv$ for our
full sample, for three cuts in $z$. At $\log\tsiiv \ga -1.2$ and $2.5
\le z \le 3.0$ (where the data are particularly good) $-0.5 \la
\tsiiii/\tsiiv \la 0.5$, corresponding to a direct upper limit of $T \la
10^{4.9}-10^{5.1}\,$K for the bulk of the gas giving rise to this \SiIV\
absorption (the lower value pertaining to the higher end of the
$\tsiiv$ range.)  The simulations give predictions that depend only
very weakly on the UVB model and are in good agreement with the data,
particularly at $2.5 \le z \le 3.0$ (however note that at $z \ga 3.0$
there is somewhat more \SiIII\ absorption predicted than observed for
high $\tsiiv$).  Concentrating on the QG model, we find $\chi^2/{\rm
d.o.f.}=80.5/114$ comparing all simulated and observed points with
$\log \tsiiv \ge -2.0$.  Fitting an offset to the median $\log
\tsiiii/\tsiiv$ ratios (as done above for $\tciv/\tsiiv$) we find
$-0.32_{-0.16}^{+0.14}\,$dex, indicating that simulations predict
somewhat too much \SiIII\ absorption overall.  We have repeated this
exercise for the 16th, 31st, 69th and 84th percentiles in order to
fit the center and width $\sigma({\rm [SiIII/SiIV]})$ of a lognormal
distribution governing scatter in $\tsiiii/\tsiiv$ beyond that present
in the simulations. The fit obtained indicates that overall the
observed $\tsiiii/\tsiiv$ ratio is lower by $0.15\pm0.06\,$dex than that for
the simulations, with $\sigma({\rm [SiIII/SiIV]})=0.07\pm0.07\,$dex.
These results imply that most of the gas is cool enough to be
consistent with the photoionization equilibrium assumed in the
simulations, and that the scatter in $\tsiiii/\tsiiv$ is similar to that in
the simulations.  But they are also consistent with a (small)
contribution by hotter gas to the observed \SiIV\ optical depths.

\section{Analysis and discussion of results}
\label{sec-discuss}

We have shown that the pixel optical depth correlations derived from
our observed QSO spectra are consistent, in detail, with spectra drawn
from a hydrodynamical simulation with: 1) an assumed carbon
metallicity distribution as derived in Paper II from measurements of
\CIV\ absorption, 2) a uniform (rescaled) UVB model taken
from~\citet{haardt01:cuba}, and 3) a constant and uniform [Si/C]
value.

We have used this success to draw inferences regarding [Si/C] for each
of several models of the UVB.  Before analyzing these inferences it is
worth discussing some effects that bear on their robustness.

\subsection{Uncertainties}
\label{sec-uncertain}

{\em UVB spectral shape:} Our inference of [Si/C] obviously depends on
  the assumed shape and evolution of the UVB: the models used here
  produce a range $0.26 \le {\rm [Si/C]} \le 1.48$.  These models,
  however, span a range of possibilities that is larger than that
  allowed by independent observations.  The harder model Q, for
  example, was found in Paper II to produce a mean carbon abundance
  that increases with decreasing density, which is probably
  unphysical.  Likewise the softer model QGS was found to imply
  increasing [C/H] with $z$, also unphysical.  The model QGS3.2, with a
  sudden transition at $z=3.2$, produced no problems in Paper II, but
  here we find that the resulting predicted jump in $\tsiiv/\tciv$ is
  disfavored by our observations.
  Our fiducial model QG appears compatible (and produces physical
  results) for all of our metallicity measurements thus far; it
  therefore seems unlikely that this UVB is very far in error. Note,
  however, that~\citet{boksen} found that they were unable to fit
  \SiII/\SiIV\ vs. \CII/\CIV\ column density ratios in their data
  (probing higher densities than our analysis) using Haardt \& Madau
  (2001) models.

{\em UVB inhomogeneity:} We have assumed that the UVB is perfectly
  uniform, which is probably unrealistic.  However, we can limit the
  non-uniformity of the hardness of the UVB because such
  non-uniformity would lead to scatter in $\tsiiv/\tciv$ even for
  uniform [Si/C], whereas our measurement of the higher percentiles in
  $\tsiiv/\tciv$ proved to be compatible with the predictions of the
  simulations for a uniform [Si/C].  

{\em Collisionally ionized gas:} The simulation we have employed does
 not contain feedback from star formation and therefore contains only
 the hot ($T\gg 10^5\,$K) gas resulting from heating by accretion
 shocks.  If the true universe has hot gas not present in our
 simulations, our inferences could be biased.  We have put a
 constraint $T\la 10^{4.9}\,$K on the temperature of the bulk of the
 gas giving rise to the \SiIV\ absorption, using the $\tsiiii/\tsiiv$
 ratio.  The simulation, however, does predict slightly higher
 $\tsiiii/\tsiiv$ overall, by $\approx 0.3\,$dex using the median and
 by $\approx 0.15\,$dex using all percentiles.  Although this might be
 attributed to some systematic difference between simulations and
 observations, it may indicate that some fraction ($\la 30\%$, since
 that is the lowest percentile we probe) of the \SiIV\ absorbing gas
 is hotter ($T > 10^5\,$K) than in the simulations. Note also that, as
 in Paper II, we can put no constraint on metals in gas that is either
 too hot ($T \gg 10^5\,$K) or too cold ($T \ll 10^4$\,K) to cause
 detectable \SiIV\ or \CIV\ absorption.

{\em Uncertainties in recombination rates:}
\citet{2000ApJ...533..106S} has pointed out that the \SiIV\ and \CIV\
dielectric recombination rates used in CLOUDY may be uncertain by up
to $\sim 0.3$\,dex, leading to uncertainties in inferred [Si/C].
Considering first \CIV, the laboratory work
of~\citet{2001ApJ...555.1027S}, with quoted uncertainties of $\sim
15\%$, finds rates $\sim 15-30\%$ higher than those used in CLOUDY
version 94.  The calculation of~\citet{2000ApJ...533..106S} suggest
that if the CLOUDY rate were $30\%$ too low, we would overestimate
[Si/C] by a factor that depends on the density of the absorbing
gas and varies from $\sim 0.0-0.07$\,dex for the range given below
in~\S~\ref{sec-dens}. For \SiIV\ the situation is worse because there
are no published laboratory experiments, so the rate could be
uncertain by $\pm0.3$\,dex~\citep{2000ApJ...533..106S}.  If the CLOUDY
rate were too high (resp. too low) by $0.3\,$dex, we would
overestimate (resp. underestimate) [Si/C] by $0.2\,$dex at the highest
gas densities we probe, by a negligible amount at the lowest densities,
and probably by $\sim 0.1\,$dex overall.

Note also that differences in the Si and C line widths, as discussed
in~\S~\ref{sec-sivhi} could lead to an uncertainty of $\sim 0.1\,$dex in
inferred [Si/C] using our method.

\subsection{Corresponding gas and column densities}
\label{sec-dens}

We have analyzed \SiIV, \CIV,\ and \HI\ pixel optical depths to draw
inferences regarding the cosmic silicon abundance.  For these to be
compared with those from theoretical or observational studies, it is useful to
estimate the gas densities and the \HI\ and \CIV\ line column
densities our measurements pertain to.

In measuring $\tsiiv/\thi$ vs. $\thi$, (Fig.~\ref{fig:sh_zcuts}), we
obtain 2$\sigma$ detections for $0.75 \la \log\thi \la 2.5$ for $2.5
\le z \le 3.5$ (where our data are best).  Using the theoretical
$\thi$-density relation given in Paper II, this range corresponds at
$z=3$ to $0.5 \la \log\delta \la 2.0$. (Note, however, that the upper
limit results from the inability to recover high-$\thi$ pixels and
does not apply to results binned in $\tciv$.)  These densities can be
converted into approximate \HI\ column densities using the theoretical
relation $N_{\rm HI} \sim 10^{15}\,{\rm
cm^{-2}}(\delta/10)^{3/2}[(1+z)/4]^{9/2}$
from~\cite{2001ApJ...559..507S}, to obtain a column density range
\begin{equation}
2\times 10^{14}\,{\rm
cm^{-2}} \la N_{\rm HI} \la 3\times 10^{16}\,{\rm cm^{-2}}.
\end{equation}

Another way to estimate the \HI\ column densities corresponding to
the \HI\ optical depth range over which we detect \SiIV\ is to use
the relation between
column density $N$ and optical depth $\tau_c$ at the center of a line
of width $b$, 
\begin{equation}
\tau_c=0.76\left ({f\lambda_0\over f_{\rm HI}\lambda_{0,{\rm HI}}
}\right )
\left({N\over{10^{13}}\,{\rm cm^{-2}}}\right)\left({b\over
10\,{\rm km\,s^{-1}}}\right)^{-1},
\end{equation}
where $f$ and $\lambda_0$ are the oscillator strength and the rest
wavelength respectively. Using $b=25~{\rm km\,s^{-1}}$ yields
$2~\times~10^{14} {\rm cm^{-2}} \la N_{\rm HI}\la 1 \times 10^{16}{\rm
cm^{-2}}$, in close agreement with our earlier estimate. We can
estimate the \CIV\ column density range in a similar way. From
Fig.~\ref{fig:sc_zcuts} we see that \SiIV\ is detected over the range
$-1.5 \la \log\tau_{\rm CIV} \la 0.0$, which corresponds to
\begin{equation}
7\times 10^{11}{\rm cm^{-2}} \la N_{\rm CIV} \la 2\times 10^{13}{\rm
  cm^{-2}}, 
\end{equation}
for a typical \CIV\ line width of $10~{\rm km\,s^{-1}}$.

Note also that our analysis pertains to significantly lower physical
densities than results obtained using the \CII/\CIV\ ratio as a
density tracer~\citep{1998AJ....115.2184S,boksen}; using Fig.~23
of~\citet{boksen}, our upper density limit corresponds roughly to
$N({\rm CII})/N({\rm CIV}) \la -1.5$.\footnote{If the relatively
high-density gas probed by~\cite{boksen} (and Songaila 1998)
lies relatively close to galaxies, it may be strongly
affected by those galaxies' soft UV radiation.}

\subsection{Comparison to Nucleosynthetic yield studies}

The small scatter in [Si/C] implies that silicon and carbon share a
common enrichment mechanism, and both the high [Si/C] and early
enrichment epoch we observe suggest enrichment by massive stars.  It
is therefore interesting to compare our [Si/C] to that predicted in
theoretical supernova yields for low-metallicity, zero metallicity,
and (more speculatively) supermassive progenitors.  The yields
of~\citet{1995ApJS..101..181W}, integrated over a Salpeter initial
mass function (IMF) with slope between -2.5 and -0.5, give production
factors (yields relative to solar metallicity) of $0.4 \la {\rm
[Si/C]} \la 0.6$ for $10^{-4} \le Z/Z_\odot \le 0.1$ progenitor
metallicities.  At $Z=0$, significantly smaller [Si/C] results:
\citet{1995ApJS..101..181W} and~\citet{2002ApJ...567..532H} give
production factors [Si/C]$\simeq 0.15$ for massive 12-40$\msol$ stars,
and [Si/C]$\sim 0.0-0.2$ for $15-30\msol$ metal-free stars are also
quoted by~\citet{2002ApJ...577..281C} from both their own calculations
and those of~\citet{2002ApJ...565..385U}.  Zero metallicity also,
however, allows (at least theoretically) for supermassive stars with
quite different yields. \citet{2002ApJ...567..532H} provide yields of
[Si/C]$\simeq 1.2$ for supermassive (140-260$\msol$)
progenitors that produce pair-instability supernovae.  Including as well
the massive ($12-40\msol$) stars with a Salpeter IMF gives
[Si/C]$\simeq 0.85$.

We may also compare our [Si/C] values to those in metal-poor stars, but here
there are a range of results, and substantial uncertainties --
particularly with regard to C -- concerning the importance of mixing
and accretion from a companion~\citep[e.g.,][]{2000A&A...356..238C}.
The lowest-metallicity (${\rm [Fe/H]} \la -2.5$) halo stars give
typically [Si/C]$\sim -0.4 -
+0.5$~\citep[e.g.,][]{2002ApJ...577..281C,2001ApJ...561.1034N,2002A&A...390..187D,1995AJ....109.2757M}.
For $-2.5 \le {\rm [Fe/H]} \le -1.5$, somewhat higher values are
apparently seen: measurements of $-0.5 \la {\rm [C/Fe]} \la 0.0$ and
$0.2 \la {\rm [Si/Fe]}\la 0.7$ (see the compilation by Norris et
al. 2001) suggest ${\rm [Si/C] \sim 0.7}$,
and~\citet{1995AJ....109.2757M} finds values of ${\rm [Si/C]}\sim
0.2-1.2$.  However,~\citet{2000A&A...356..238C} find higher [C/Fe] of
$-0.2 \la {\rm [C/Fe]} \la 0.3$ when using only unmixed stars,
implying somewhat smaller ${\rm [Si/C]}\sim 0.4$.

If we assume that our fiducial UVB model (and assumed recombination
rates; see~\S\ref{sec-uncertain}) are correct, then the inferred
[Si/C]$=0.77\pm 0.05$ has interesting implications.  It is
substantially higher than the [Si/C] predicted from $\la 40\msol$
Population III stars with a standard IMF, and it is somewhat ($\sim
0.2\,$dex) higher than abundance ratios in low-metallicity stars, and
predicted yields for massive Population II stars.  If the latter discrepancy
is taken seriously, it could be remedied by including some
supermassive (pair-instability) supernovae as modeled
by~\citet{2002ApJ...567..532H}.\footnote{That supermassive stars might
be necessary to explain our observed [Si/C] value is particularly
interesting in light of the appreciable fraction of all cosmic metal
production at $z \ga 3$ that may be inferred to lie in the Ly$\alpha$
forest (see \S~\ref{sec-abun}).}

If, instead, we {\em assume} that [Si/C] should be $\sim 0.5$ on the basis of
standard nucleosynthesis yields, then we infer that the UVB must be
softer (at bith low- and high-$z$) than our fiducial model
QG. However, it should be noted that a UVB as soft as model QGS, for
which we find [Si/C]$=0.26$, was demonstrated to produce unphysical
results in Paper II.

\subsection{Implications for cosmic abundances}
\label{sec-abun}

In \S~\ref{sec-resrel} we measured [Si/C] values for several UVB
models by comparing observations of $\tsiiv/\tciv$ to simulations
generated using a given UVB model and a distribution of [C/H] as
determined in Paper II (\S~7) for that UVB. Combining these fits (See
Table~\ref{tbl:fits}) with the results of Paper II enables us to
assess the overall cosmic abundance of silicon.  For these
calculations we assume that our derived [Si/C] is uniform and
constant, but we include the uncertainty in its determination.

In Paper II we combined the median [C/H]$(\delta, z)$ with the width
$\sigma([{\rm C/H}])(\delta,z)$ of the lognormal probability
distribution of [C/H] for $-0.5 \le \log\delta \le 2.0$ to determine
the mean C abundance versus $\delta$.  This was then integrated over the
mass-weighted probability distribution $\delta$ (obtained from our
hydrodynamical simulation) to compute the contribution by gas in this
density range to the overall mean cosmic [C/H].  Assuming that [Si/C]
is constant over this density range\footnote{Note that this is an
extrapolation beyond the range over which we have {\em measured}
[Si/C].} we obtain, for our fiducial UVB model QG, [Si/H]$= -2.03 \pm
0.14$, corresponding to
\begin{equation}
\Omega_{\rm Si,IGM} \simeq 3.4\times10^{-7} 10^{[{\rm Si}/{\rm H}]+2.0}\left ({\Omega_b
\over 0.045}\right ),
\end{equation}
with no evidence for evolution.  Extrapolating our [C/H] and [Si/C]
results to the full density range of the simulation would yield values
$\approx 0.2$\,dex higher.  For a harder UVB, the inferred cosmic Si
abundance would be significantly higher, as both the inferred [C/H]
and [Si/C] increase with the UVB hardness.  For our quasar-only model
Q, we obtain a contribution of [Si/H]$\approx -0.8$, or $\Omega_{\rm
Si} \approx 5\times10^{-6}$.  For comparison,
\citet{2001ApJ...561L.153S}\footnote{As in Paper II, we have corrected
Songaila's value, which assumes an $\Omega_m=1.0$ cosmology to the
$\Omega_m=0.3$ cosmology adopted here.} found, using a direct sum of
\SiIV\ lines, that $\Omega_{\rm SiIV} = (0.5-2)\times 10^{-8}$.  This
indicates that although Songaila's results provide a useful lower
limit, an ionization correction of more than 10 (or even much larger for
harder backgrounds) must be made to translate $\Omega_{\rm SiIV}$ into
an estimate of the true cosmic silicon abundance. The large ionization
correction implies that the evolution of $\Omega_{\rm SiIV}$ does not
by itself provide interesting constraints on the evolution of
$\Omega_{\rm Si}$.

It is interesting to compare our derived metallicities to the metal
production expected of high-$z$ galaxies and contained in other known
types of objects.  Integration of the observed cosmic star formation
rate indicates that $\Omega_*(z=2.5) \approx 0.001$, a quantity
sufficient to supply a contribution of 1/30 solar
metallicity~\citep[see, e.g.,][]{pettinirev}.  Thus if Si is used as a
tracer of metals, then the metals observed in the Ly$\alpha$ forest
would represent $\approx 30-50\%$ (for the QG UVB, depending on
whether we extrapolate to high-densities) of all metals produced by
$z=2.5$.  If the Q UVB were to hold, the IGM would hold $\approx
5-8$ times the expected metal production, which is probably yet
another indication that the Q UVB is unrealistically hard.  Another,
similar approach is to compare our derived $\Omega_{\rm Si}$ to the
total amount of Si locked in stars:
\begin{equation}
\Omega_{\rm Si,*}\simeq7.6\times10^{-7}\left({\Omega_*\over
  0.001}\right){Z_{\rm Si}\over Z_{\rm Si,\odot}},
\end{equation}
where $Z_{{\rm Si}}$ is the mean silicon metallicity in those stars,
and $\Omega_*\approx 0.001$ is the estimate given
by~\citet{pettinirev} for the stellar density at $z=2.5$.  Thus, even
if all stars at $z\approx 2.5$ were to carry solar abundance of
silicon, they would contain $\sim 60-70\%$ (for QG) or $\sim 8-12\%$
(for Q) of the cosmic silicon; the rest would be stored in the IGM.

It is also worth noting that, if silicon is used as the metallicity
tracer and our fiducial UVB is adopted, the Ly$\alpha$ forest contains
about twice the cosmic metal mass provided by damped Lyman absorbers,
as estimated by~\citet{pettinirev}.

\section{Conclusions}
\label{sec-conc} 

We have studied the relative abundance of silicon in the IGM by analyzing
\SiIV, \CIV, and \HI\ pixel optical depths derived from a set of 
high-quality VLT and Keck spectra of 19 QSOs at $1.5\la z \la 4.5$, and
comparing them to realistic, synthetic spectra drawn from a
hydrodynamical simulation to which metals have been added.  Our fiducial
model employs the ionizing background model (``QG'') taken from Haardt
\& Madau (2001) for quasars and galaxies (rescaled to reproduce the
observed mean Ly$\alpha$ absorption), and assumes a carbon abundance
as derived in 
Paper II: at a given density $\delta$ and redshift $z$, [C/H] has a
lognormal probability distribution centered on
$-3.47+0.65(\log\delta-0.5)$ and of width $0.70\,$dex.  The main
conclusions of this analysis are as follows:
 
\begin{itemize}
\item{For our fiducial model, the median optical depth ratios
$\tsiiv/\tciv$ for $-1.5\la \log \tciv \la 0$ are reproduced well as a
function of $z$ and $\tciv$ by the simulations, if we take
[Si/C]=$0.77\pm0.05$, uniformly and at all times.  These measurement
pertain to gas (over)densities of $\delta > 3$, or column densities
$N_{\rm HI}\ga 2\times 10^{14}\,{\rm cm^{-2}}$ and $N_{\rm CIV}
\approx 7\times 10^{11}-2\times 10^{13}\,{\rm cm^{-2}}$ at $z=3$.}
\item{We find a strong correlation between $\tsiiv/\tciv$ and $\tciv$
exhibited by the data that are reproduced by the simulations. This
indicates that evolution in $\tsiiv/\tciv$ is best assessed using
small cuts in $\tciv$. Our results agree very well with simulations
using a smoothly evolving UVB\, and show no significant evolution in
$\tsiiv/\tciv$ (except perhaps a slight rise in $\tsiiv/\tciv$ with
$z$ for $\log\tciv \ga -0.5$). The lack of evolution is consistent
with the results of~\citet{boksen} and Kim et al. (2002), but
not those of~\citet{1998AJ....115.2184S}.}
\item{The [Si/C] value inferred from $\tsiiv/\tciv$ depends on the
assumed UVB. A harder UVB model Q (using only the contribution of
quasars to the UVB) gives a much higher ratio,
[Si/C]=$1.48^{+0.05}_{-0.06}$. A model ``QGS3.2'' in which the UVB is
decreased in intensity by 1\,dex blueward of 4 Ryd at $z > 3.2$
(crudely simulating incomplete \HeII\ reionization) gives a lower
ratio $0.46^{+0.10}_{-0.08}$ (or $0.32^{+0.08}_{-0.13}$ using the
$z>3$ data alone), but gives too much evolution in $\tsiiv/\tciv$.}
\item{Subdividing the sample by $z$ and gas density $\delta$, we see
no evidence for evolution in [Si/C] for any of our UVBs (except
QGS3.2, which gives significantly different results at low- and
high-$z$).  The low-$\delta$ and high-$\delta$ [Si/C] inferences are
discrepant by $1.4\sigma$ for our fiducial UVB and by $3.8\sigma$ for
the quasar-only UVB.  This indicates either that [Si/C] increases with
decreasing density, or that our fiducial UVB may be slightly too hard
(and that the QSO-only UVB is far too hard).}
\item{While the dominant uncertainty in [Si/C] comes from the shape of
the UVB, an additional uncertainty of up to $\sim \pm 0.1$~dex in our
inferences (and those of other studies) could result from
uncertainties in the assumed \SiIV\ dielectric recombination rates.
There may also be a systematic uncertainty of up to $\sim 0.1\,$dex
resulting from the different thermal widths of C and Si.}
\item{Evaluating [Si/C] for optical depth percentiles higher than the
median gives no evidence for additional scatter, either in Si/C or in
the hardness of the UV background (and hence the ionization
correction). The width of a lognormal distribution\footnote{Note that
because we can only obtain percentiles in $\tsiiv/\tciv$ near and
above the median, our data only constrains the upper {\em half} of the
distribution.}  of [Si/C] is constrained to be much smaller
than that of [C/H].}
\item{Analysis of $\tsiiv/\thi$ also gives information on [Si/H] and
[Si/C] but is subject to large systematic effects because of the
significantly different thermal width of Si and H.  If the data is
smoothed to minimize this difference, we can roughly reproduce the
inferences based on $\tsiiv/\tciv$.}
\item{The measured ratios $\tau_{\rm SiIII}/\tsiiv$ provide an upper
limit $T < 10^{4.9}\,$K on the temperature of the bulk of the gas
responsible for the \SiIV\ absorption.  The measured $\tau_{\rm
SiIII}/\tsiiv$ versus $\tsiiv$ and $z$ are roughly in accord with, but
$\sim 0.15\,$dex lower than, the predictions of the simulations.
This may be an indication that a small fraction of the observed gas is
at higher temperature than in the simulations.}
\item{Our inferred [Si/C] is $\sim 0.2\,$dex higher than that
predicted by Population II, Type II supernova yield calculations based on a
standard IMF up to $\sim 40\msol$, and that observed in metal-poor
stars, and much higher than predicted for the yields of massive $(M<
40\msol)$ Population III stars.  High [Si/C] values can, however, be
obtained from an IMF that includes supermassive Population III stars
exploding as pair-instability supernovae.  Alternatively, we could
conclude that our fiducial UVB is too hard; however, a UVB model
significantly softer than our fiducial one leads to unphysical results
such as negatively evolving C metallicity (see Paper II).}
\item{Combining our measured [Si/C] with
the measurements of [C/H] of paper II, we find that the Ly$\alpha$
forest contributed [Si/H]$=-2.0$ to the global silicon abundance for
the QG UVB, or $\Omega_{\rm Si} \simeq 3.2\times10^{-7}$.  
This would constitute $\approx 30-50\%$ of the expected Si production by
$z=2.5$ as estimated by~\cite{pettinirev}.}
\end{itemize}

\acknowledgements We are grateful to the ESO Archive for their
efficient work. Without their help this work would not have been
possible. AA and JS gratefully acknowledge support from the W.~M.~Keck
foundation, and JS acknowledges support from NSF grant PHY-0070928.
TT thanks PPARC for the award of an Advanced Fellowship. WLWS
acknowledges support from NSF Grant AST-0206067. MR is grateful to the
NSF for grant AST-00-98492 and to NASA for grant AR 90213.01-A. We
thank Jason Prochaska and Stan Woosley for helpful conversations. This
work has been conducted with partial support by the Research Training
Network "The Physics of the Intergalactic Medium" set up by the
European Community under the contract HPRN-CT2000-00126 RG29185 and by
ASI through contract ARS-98-226. Research has been conducted in cooperation
with Silicon Graphics/Cray Research utilising the Origin 2000
supercomputer at DAMTP, Cambridge.


\begin{thebibliography}{}

\bibitem[Paper(I)]{paper1} Aguirre, 
A., Schaye, J., \& Theuns, T.\ 2002, \apj, 576, 1  (Paper I)

\bibitem[Anders \& Grevesse(1989)]{1989GeCoA..53..197A} Anders, E.~\& 
Grevesse, N.\ 1989, \gca, 53, 197 

\bibitem[Aracil et al.(2003)]{aracil} Aracil, B., Petitjean, P.,
Pichon, C., \& Bergeron, J.\ 2003, \apjl, submitted; astro-ph/0307506

\bibitem[Boksenberg et al.(2003)]{boksen} Boksenberg, A.,
Sargent, W.L.W., \& Rauch, M.\ 2003, \apjs, submitted; astro-ph/0307557

\bibitem[Carretta, Gratton, \& Sneden(2000)]{2000A&A...356..238C} Carretta, 
E., Gratton, R.~G., \& Sneden, C.\ 2000, \aap, 356, 238 

\bibitem[Chieffi \& Limongi(2002)]{2002ApJ...577..281C} Chieffi, A.~\& 
Limongi, M.\ 2002, \apj, 577, 281 

\bibitem[Cowie \& Songaila(1998)]{1998Natur.394...44C} Cowie, L. L. \& 
Songaila, A. 1998, \nat, 394, 44 

\bibitem[Cowie et al.(1995)]{1995AJ....109.1522C} Cowie, 
L.~L., Songaila, A., Kim, T., \& Hu, E.~M.\ 1995, \aj, 109, 1522 

\bibitem[Croft et al.(1997)]{1997ApJ...488..532C} 
Croft, R.~A.~C., Weinberg, D.~H., Katz, N., \& Hernquist, L.\ 1997, \apj, 
488, 532 

\bibitem[Dav{\' e} et al.(1998)]{1998ApJ...509..661D} Dav{\' e}, R.,
 Hellsten, U., Hernquist, L., Katz, N., \& Weinberg, D.~H.\ 1998, \apj, 509,

\bibitem[Depagne et al.(2002)]{2002A&A...390..187D} Depagne, E.~et al.\ 
2002, \aap, 390, 187 

\bibitem[D'Odorico et al.(2000)]{2000SPIE.4005..121D} D'Odorico, S., 
Cristiani, S., Dekker, H., Hill, V., Kaufer, A., Kim, T., \& Primas,
F.\ 2000, \procspie, 4005, 121 

\bibitem[Ellison et al.(1999)]{1999ApJ...520..456E} Ellison, S.~L., Lewis, 
G.~F., Pettini, M., Chaffee, F.~H., \& Irwin, M.~J.\ 1999, \apj, 520, 456 

\bibitem[Ellison et al.(2000)]{2000AJ....120.1175E} 
Ellison, S.~L., Songaila, A., Schaye, J., \& Pettini, M.\ 2000, \aj, 120, 
1175 

\bibitem[Ferland(2000)]{2000RMxAC...9..153F}
Ferland, G.~J.\ 2000, Revista Mexicana de Astronomia y Astrofisica
Conference Series, 9, 153

\bibitem[Ferland et al.(1998)]{1998PASP..110..761F}
Ferland, G.~J., Korista, K.~T., Verner, D.~A., Ferguson, J.~W.,
Kingdon, J.~B., \& Verner, E.~M.\ 1998, \pasp, 110, 761

\bibitem[Haardt \& Madau(2001)]{haardt01:cuba}
Haardt, F.~\& Madau, P. 2001, to be published in the proceedings of
XXXVI Rencontres de Moriond, astro-ph/0106018

\bibitem[Heger \& Woosley(2002)]{2002ApJ...567..532H} Heger, A.~\& Woosley, 
S.~E.\ 2002, \apj, 567, 532 

\bibitem[Kim, Cristiani, \& D'Odorico(2002)]{2002A&A...383..747K} Kim, 
T.-S., Cristiani, S., \& D'Odorico, S.\ 2002, \aap, 383, 747 

\bibitem[McWilliam et al.(1995)]{1995AJ....109.2757M} 
McWilliam, A., Preston, G.~W., Sneden, C., \& Searle, L.\ 1995, \aj, 109, 
2757

\bibitem[Norris, Ryan, \& Beers(2001)]{2001ApJ...561.1034N} Norris, J.~E., 
Ryan, S.~G., \& Beers, T.~C.\ 2001, \apj, 561, 1034 

\bibitem[Pieri \& Haehnelt(2003)]{pieri} Pieri, M., \& Haehnelt, M.\
2003, \mnras, submitted; astro-ph/0308003

\bibitem[Pettini (2003)]{pettinirev} Pettini, M. 2003, astro-ph/0303272

\bibitem[Ryan, Norris, \& Beers(1996)]{1996ApJ...471..254R} Ryan, S.~G., 
Norris, J.~E., \& Beers, T.~C.\ 1996, \apj, 471, 254 

\bibitem[Savin(2000)]{2000ApJ...533..106S} Savin, D.~W.\ 2000, \apj, 533, 
106 

\bibitem[Schaye(2001)]{2001ApJ...559..507S} Schaye, J.\ 2001, \apj, 559, 
507 

\bibitem[Paper(II)]{paper2} Schaye, J., Aguirre, A., Kim, T., Theuns,
T., Rauch, M., \& Sargent, W.L.W.\ 2003, \apj, 596, 768


\bibitem[Schaye et al.(2000a)]{2000ApJ...541L...1S} Schaye, 
J., Rauch, M., Sargent, W.~L.~W., \& Kim, T.\ 2000a, \apjl, 541, L1 

\bibitem[Schaye et al.(1999)]{1999MNRAS.310...57S} 
Schaye, J., Theuns, T., Leonard, A., \& Efstathiou, G.\ 1999, \mnras, 310, 
57 

\bibitem[Schaye et al.(2000b)]{2000MNRAS.318..817S} Schaye, J., Theuns, T., 
Rauch, M., Efstathiou, G., \& Sargent, W.~L.~W.\ 2000b, \mnras, 318, 817 

\bibitem[Schippers et al.(2001)]{2001ApJ...555.1027S} Schippers, S., M{\" 
u}ller, A., Gwinner, G., Linkemann, J., Saghiri, A.~A., \& Wolf, A.\ 2001, 
\apj, 555, 1027 

\bibitem[Songaila(1998)]{1998AJ....115.2184S} Songaila, A.\ 1998, \aj, 115, 
2184 

\bibitem[Songaila(2001)]{2001ApJ...561L.153S} Songaila, A.\ 2001, \apjl, 
561, L153 

\bibitem[Songaila \& Cowie (1996)]{1996AJ....112..335S} Songaila, A.  \& 
Cowie, L. L. 1996, \aj, 112, 335 

\bibitem[Theuns et al.(2002)]{2002ApJ...578L...5T} Theuns, T., Viel, M., 
Kay, S., Schaye, J., Carswell, R.~F., \& Tzanavaris, P.\ 2002, \apjl, 578, 
L5 

\bibitem[Umeda \& Nomoto(2002)]{2002ApJ...565..385U} Umeda, H.~\& Nomoto, 
K.\ 2002, \apj, 565, 385 

\bibitem[Vogt et al.(1994)]{1994SPIE.2198..362V} Vogt, S.~S.~et al.\ 1994, 
\procspie, 2198, 362 

\bibitem[Woosley \& Weaver(1995)]{1995ApJS..101..181W} Woosley, S.~E.~\& 
Weaver, T.~A.\ 1995, \apjs, 101, 181 

\end{thebibliography}
\end{document}